\documentclass[acmtog]{acmart} 

\usepackage{booktabs}

\setcitestyle{nosort,square} %


\usepackage{hyperref}
\usepackage{amsmath}
\usepackage{graphicx}
\usepackage{comment}
\usepackage{kbordermatrix}
\usepackage{multirow,bigdelim}

\usepackage{array}
\usepackage{wrapfig}

\usepackage{csvsimple}
\usepackage{adjustbox}
\usepackage[percent]{overpic}
\usepackage{makecell}
\usepackage{float}

\usepackage{blindtext}
\usepackage{float}
\usepackage{placeins}

\newcolumntype{C}[1]{>{\centering\let\newline\\\arraybackslash\hspace{0pt}}m{#1}}

\newif\ifdraft
\draftfalse

\ifdraft
\newcommand{\dcc}[1]{{\color{red}[\textbf{DC:} #1]}}
\newcommand{\rhc}[1]{{\color{orange}[\textbf{RH:} #1]}}
\newcommand{\gmc}[1]{{\color{blue}[\textbf{GM:} #1]}}
\newcommand{\rgc}[1]{{\color{olive}[\textbf{RG:} #1]}}

\newcommand{\dc}[1]{{\color{red}#1}}
\newcommand{\rh}[1]{{\color{orange}#1}}
\newcommand{\gm}[1]{{\color{blue}#1}}
\newcommand{\rg}[1]{{\color{olive}#1}}

\else
\newcommand{\dcc}[1]{}
\newcommand{\rhc}[1]{}
\newcommand{\rgc}[1]{}
\newcommand{\gmc}[1]{}
\newcommand{\dc}[1]{{\color{black}#1}}
\newcommand{\rh}[1]{{\color{black}#1}}
\newcommand{\rg}[1]{{\color{black}#1}}
\newcommand{\gm}[1]{{\color{black}#1}}

\fi

\newcommand{\ourmethod}{self-sample net}

\newcommand{\rev}[1]{{\color{black}#1}}
\newcommand{\revTwo}[1]{{\color{blue}#1}}

\usepackage{pgfplots}
\pgfplotsset{compat=newest}
\usepgfplotslibrary{groupplots}
\usepgfplotslibrary{dateplot}

\usepackage[bottom]{footmisc}
\acmJournal{TOG}

\setcopyright{acmcopyright}
\acmJournal{TOG}
\acmYear{2021} \acmVolume{40} \acmNumber{5} \acmArticle{191} \acmMonth{10} \acmPrice{15.00}\acmDOI{10.1145/3470645}

\begin{document}

\title{Self-Sampling for Neural Point Cloud Consolidation}

\author{Gal Metzer}
\affiliation{\institution{Tel Aviv University}}

\author{Rana Hanocka}
\affiliation{\institution{Tel Aviv University}}

\author{Raja Giryes}
\affiliation{\institution{Tel Aviv University}}

\author{Daniel Cohen-Or}
\affiliation{\institution{Tel Aviv University}}

\begin{abstract}

We introduce a novel technique for neural point cloud consolidation which learns from only the input point cloud. 
Unlike other point upsampling methods which analyze shapes via local patches, in this work, we learn from global subsets. 
We repeatedly \textit{self-sample} the input point cloud with global subsets that are used to train a deep neural network. Specifically, we define source and target subsets according to the desired consolidation criteria (e.g., generating sharp points or points in sparse regions). The network learns a mapping from source to target subsets, and implicitly learns to consolidate the point cloud.
During inference, the network is fed with random subsets of points from the input, which it displaces to synthesize a consolidated point set. We leverage the inductive bias of neural networks to eliminate noise and outliers, a notoriously difficult problem in point cloud consolidation. The \textit{shared} weights of the network are optimized over the entire shape, learning non-local statistics and exploiting the recurrence of local-scale geometries. Specifically, the network encodes the distribution of the underlying shape surface within a fixed set of local kernels, which results in the \emph{best} explanation of the underlying shape surface. We demonstrate the ability to consolidate point sets from a variety of shapes, while eliminating outliers and noise.

\end{abstract}

\maketitle

\section{Introduction}
Point clouds are a popular representation of 3D shapes. Their simplicity and direct relation to 3D scanners make them a practical candidate for various applications in computer graphics. 
Point clouds acquired through scanning devices are a sparse and noisy sampling of the underlying 3D surface, which often contain outliers or missing regions.
Typically, such point sets lack orientation information (\emph{i.e.,} unoriented normals), which crucially indicates inside/outside information.
Moreover, point sets impose innate technical challenges since they are unstructured, irregular, unordered and lack non-local information.
Yet, downstream applications and graphics pipelines often demand \emph{pristine} point sets (\emph{e.g.,} oriented normals, uniformly sampled, sharp-feature points and noise/outlier-free).
\begin{figure}
    \centering
    \newcommand{\pl}{-4.5}
    \includegraphics[width=\columnwidth]{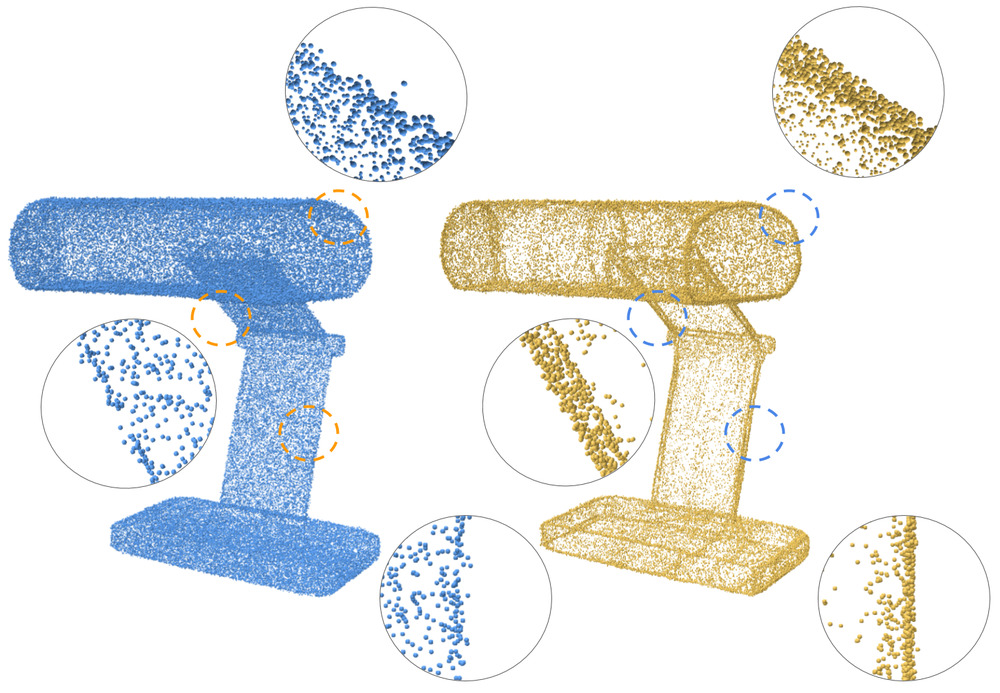}
    \begin{overpic}[width=\columnwidth]{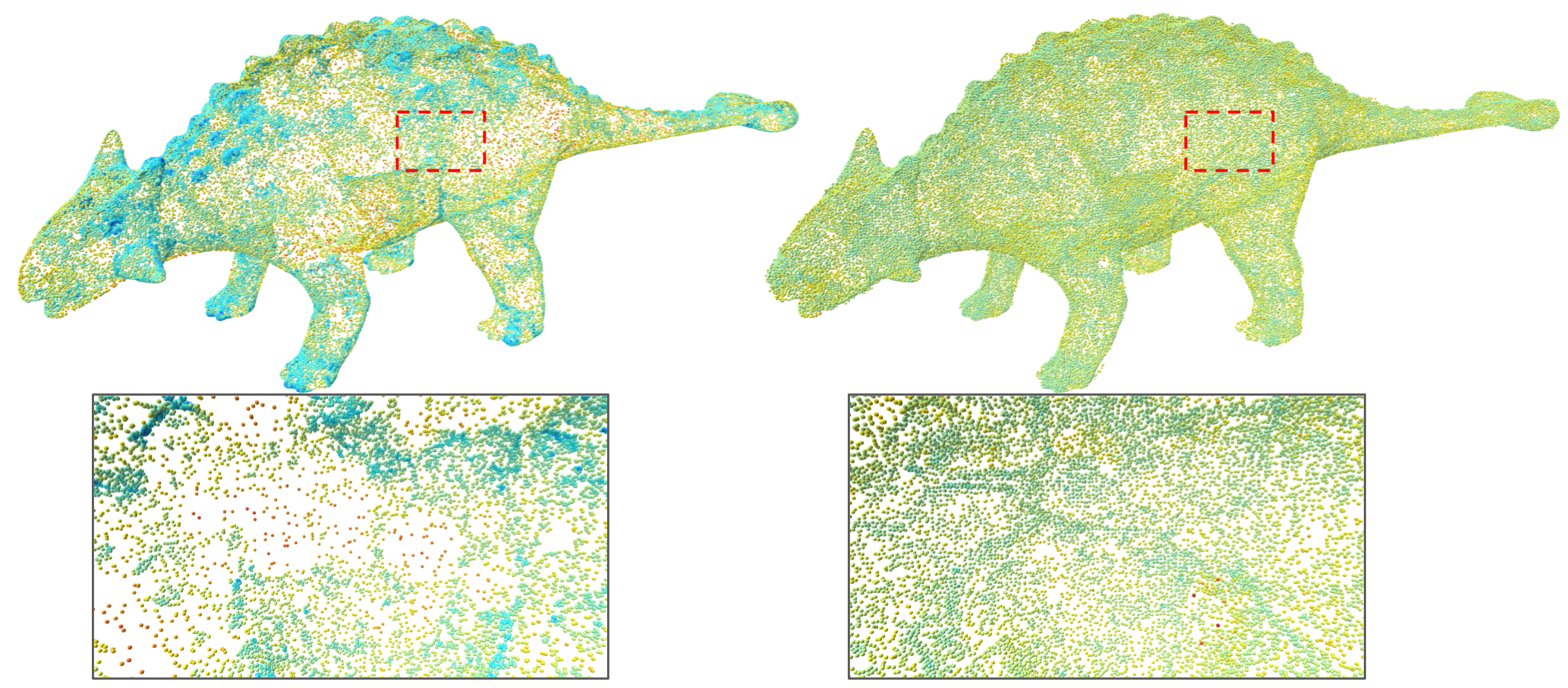}
    \put(19,  \pl){\textcolor{black}{Input}}
    \put(60, \pl){\textcolor{black}{Consolidated}}
    \end{overpic}
    \caption{Given an input point cloud (left), we generate a consolidated point cloud (right) \rh{using global self-sampling}. Our technique generates novel points on sharp features (top) and a uniformly distributed point set from sparsely sampled regions (bottom).}
    \label{fig:teaser}
\end{figure}

A variety of methods propose consolidation techniques for unorganized points sets which may contain noise, outliers, and sparsely sampled regions \cite{alexa2001point,berger2013benchmark, berger2017survey}. This consolidation process includes adding points on sharp-features and inside sparsely sampled regions as well as removing noise and outliers. In particular, a fundamental challenge in the point cloud consolidation is the representation or generation of points with sharp features or other sparsely sampled regions. Scanners struggle to uniformly sample the entire shape surface, and accurately sample sharp regions. In particular, sharp regions on continuous surfaces contain high frequencies that cannot be faithfully reconstructed using finite and discrete samples. Sharp features on 3D surfaces (\emph{e.g.,} corners and edges) are sparse, yet they are a crucial cue in representing the underlying shape. Thus, special attention is required to represent and reconstruct sharp features from imperfect data \cite{FleishmanRobust}. Furthermore, in general, redistributing points into regions with low density samples is important for understanding the underlying shape surface\rev{, and especially useful for downstream applications that use KNN (\textit{e.g.,} point cloud laplacian and normal estimation).}

Traditionally, consolidation approaches define a prior, such as piecewise smoothness, which characterizes typical 3D shapes \cite{lipman2007parameterization, huang2013edge, xiong2014robust}. Recently, \citet{yu2018ec} proposed EC-NET, a deep learning technique for edge-aware point cloud consolidation, where the prior is learned through training examples. However, these examples are created by manually annotated sharp edges for supervision.

\rh{In this work, we present a deep learning technique for point cloud consolidation using \emph{only} the single given input point cloud. Instead of learning from a collection of manually annotated shapes, we leverage the inherent self-similarity present within a single shape. We repeatedly \textit{self-sample} the input point cloud with global subsets that are used to train a deep neural network. Specifically, we define source and target subsets according to the desired consolidation criteria (e.g., generating sharp points or points in sparse regions). The network learns a mapping from source to target subsets, and implicitly learns to consolidate the point cloud.}

\textbf{Self-prior.}
Hanocka et al. \shortcite{Hanocka2020p2m} coined the self-similarity learning \textit{self-prior}, which we adopt here to indicate the innate prior learned from a single input shape. Learning internal statistics from a single image~\cite{shocher2018zero, shaham2019singan}, or on single 3D shape \cite{williams2019deep, Hanocka2020p2m, Liu:Subdivision:2020}, has demonstrated intriguing promise. In particular, self-prior learning offers considerable advantages compared to the alternate dataset-driven paradigm. For one, it does not require a supervised training dataset that sufficiently simulates the anticipated acquisition process and shape characteristics during \emph{inference} time. Moreover, a network trained on a dataset can only allocate finite capacity to one particular shape. On the other hand, in the self-prior paradigm, the entire network capacity is solely devoted to \emph{understanding} one particular shape. 
Given an input point cloud, the network implicitly learns the underlying surface and consolidation criteria, which are encoded in the network weights.

\textbf{Global Analysis.}
Unlike other point upsampling methods which analyze shapes via local patches, in this work, we learn from global subsets. 
\rh{These subsets are \textit{global} in the sense that they are sampled from the entire point cloud, which is different than existing techniques that learn on local surface patches.}
Training on subsets from the single input enables an effective technique for point cloud consolidation, which is attributed to the weight-sharing network structure. The weights are shared globally, which encode the distribution of the underlying shape surface within a fixed set of local kernels. The finite network capacity forces neurons to encode the \textit{best} explanation of the shape, which essentially prevents overfitting to noise/outliers. The inductive bias of neural networks is remarkably effective at eliminating noise and outliers, a notoriously difficult problem in point cloud consolidation.
The premise of this work is that shapes are \emph{not} random, they contain strong self-correlation across multiple scales, and can even contain recurring geometric structures (\emph{e.g.,} see Figure~\ref{fig:teaser}). The key is the internal network structure, which inherently models global recurring and correlated structures, and therefore, ignores outliers and noise.

\begin{figure}
    \centering
    \includegraphics[width=\columnwidth]{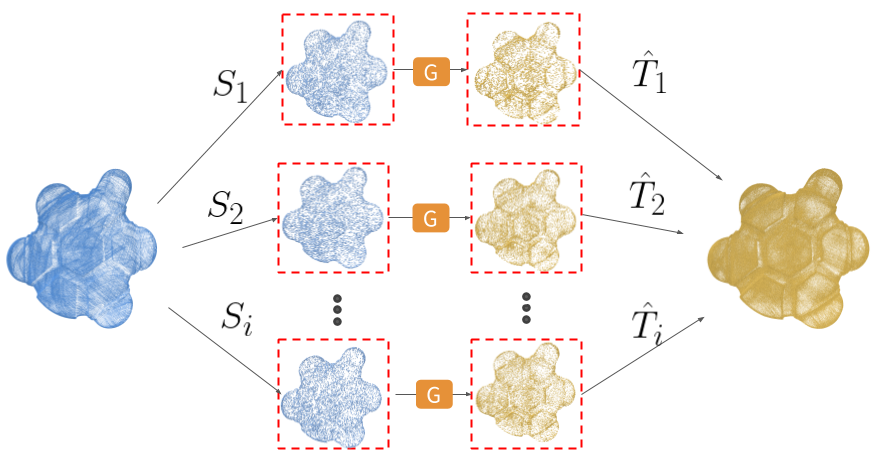}
    \caption{\rh{Inference overview. Random subsets $S_i$ are repeatedly passed through the trained network G to produce different consolidated subsets $\hat{T}_i$, which are aggregated to produce the final result.}}
    \label{fig:inference}
\end{figure}

{\bf Self-sampling.}  We train the network using self-supervision, by repeatedly selecting different global subsets of points from the input point cloud, which are used to \emph{implicitly} learn the the desired consolidation criteria. We sub-sample source and target subsets from the input point cloud according to the desired consolidation criteria (e.g., generating sharp points or points in sparse regions). The sub-sampler re-balances the sampled subsets so that target subsets contain more points than their relative prevalence in the input shape. \rh{Our network learns a mapping from source to target subsets (\emph{e.g.,}, from a set representing flat regions to a set representing sharp regions).} Specifically, the network learns to generate offsets, which are added to the input source points such that they resemble a target subset. During inference, the network is fed with random subsets of points from the input, which it displaces to synthesize a consolidated point set (Figure~\ref{fig:inference}). This process is repeated to obtain an arbitrarily large set of consolidated points with an emphasis on the consolidation criteria (\emph{e.g.,} sharp feature points or points in sparsely sampled regions).

\textbf{Consolidation Criteria.}
We use a simple criterion to classify points into two distinct groups, which is a loose approximation for the desired consolidation property. For example, we consider points with high curvature to be roughly sharp, and points with many nearby points to be roughly high-density. This approximate definition of the desired criteria is sufficient, since networks are good at regressing to the mean, understanding the core modes of the data and ignoring outliers. Obtaining the desired consolidation criteria via a rough approximation of it, can be viewed as a type of bootstrapping. This simple formulation works surprisingly well, even in the presence of noise,
since the inductive bias of neural networks acts a powerful prior, which can be used to effectively consolidate point clouds.

We demonstrate results on a variety of shapes with varying amounts of sharp features and sparsely sampled regions. We show results on point sets with noise, non-uniform density and real scanned data without any normal information. Moreover, we demonstrate the inherent denoising capabilities, as well as demonstrate the value of our consolidation in downstream tasks such as normal estimation, surface reconstruction \rh{and 3D printing}. 
In particular, we are able to accurately generate an arbitrarily large amount of points near sharp features or in sparsely sampled regions, while eliminating outliers and noise.
\begin{figure*}[h]
    \centering
    \includegraphics[width=16cm]{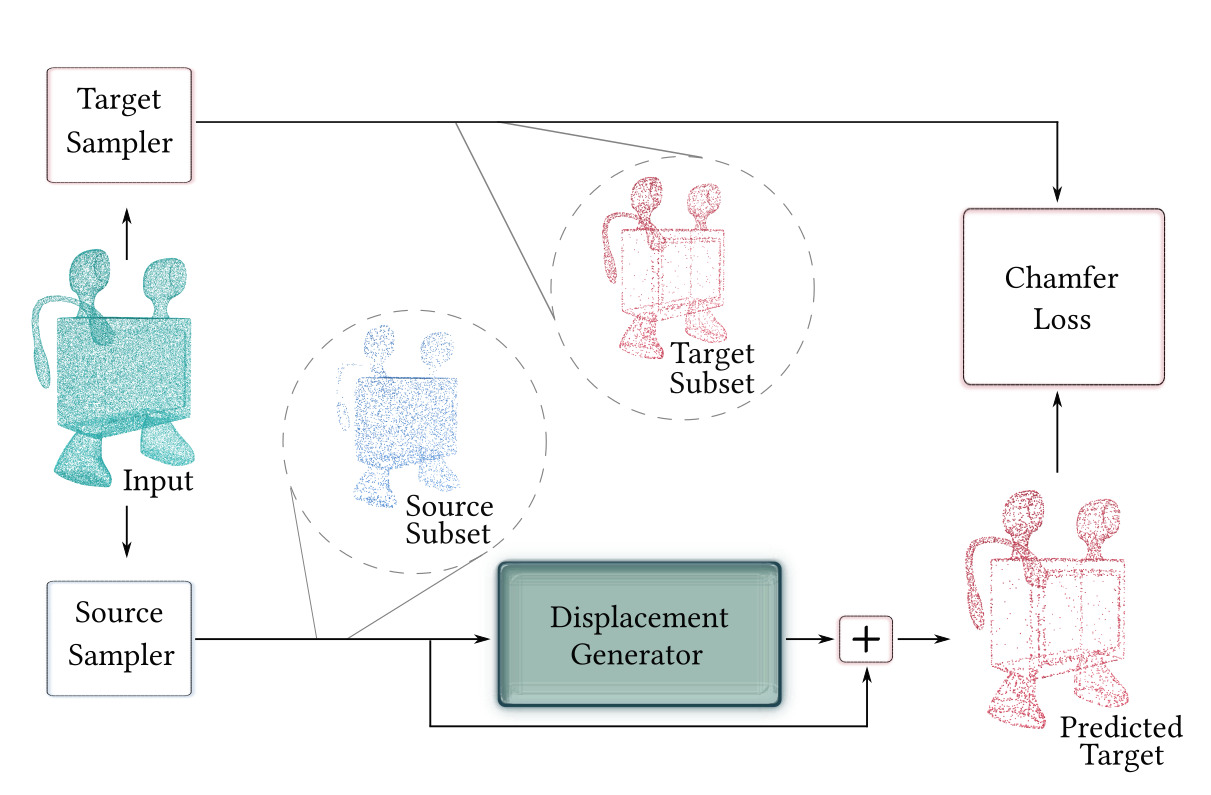}
    \caption{Training overview. We repeatedly sample \textit{source} and \textit{target} subsets from the input point cloud, where target subsets contain mainly points with the desired consolidation criterion (\emph{i.e.,} sharp). Then, the network learns displacement offsets which are added to the source subset, resulting in the predicted target. The reconstruction loss uses a bidirectional Chamfer distance between the predicted target and target subset. In each training iteration, new source and target subsets are re-sampled from the input point cloud.
    }
    \label{fig:overview}
\end{figure*}
\section{Related Work}
In recent decades there was a wealth of works for re-sampling and consolidating point clouds.
The early work of Alexa et al. \shortcite{alexa2001point} used a moving least square (MLS) operator to locally define a surface, allowing up and down sampling points on the surface. Lipman et al. \shortcite {lipman2007parameterization} presented a locally optimal projection operator (LOP) to re-sample new points that are fairly distributed across the surface. Huang et al. \shortcite{huang2009consolidation} further improved this technique by adding an extra density weights term to the cost function, allowing dealing with non-uniform point distributions.
These works showed impressive results on scanned and noisy data, but assume a smooth surface with no sharp features. To deal with piece-wise smooth surfaces that have sharp features, Haung et al. \shortcite{huang2013edge} presented a re-sampling technique that explicitly re-samples sharp features. Their technique built upon the LOP operator and approximated normals iteratively \rg{for introducing} more points along the missing or sparsely sampled edge\rg{s} using the normals in the vicinity of the sharp feature. 

Recently, with the emergence of neural networks new methods for point cloud re-sampling~\cite{yu2018pu, yifan2019patch, li2019pugan} have been introduced. 
These methods build upon the ability of neural networks to learn from data.
As well as operate on the \emph{patch-level}, which uses pairs of noisy / incomplete patches and their corresponding high-density surface sampled patches for supervision.  
\rg{Note that} processing point clouds is challenging since they are irregular and unordered, making it difficult to adopt the successful image-based CNN paradigm that assumes a regular grid structure~\cite{qi2017pointnet,qi2017pointnet++,li2018pointcnn,wang2019dynamic}. 

To up-sample a point cloud, while respecting its sharp features, EC-Net~\cite{yu2018ec} learns from a dataset of meshes. The technique is trained to identify points that are on or close to edges and allows generating edge points during up-sampling. Their method, like the above up-sampling methods, are trained on large supervised datasets. 
Therefore, their performance is highly dependent on how well the training data accurately models the expected properties of the shapes during test time (\emph{i.e.,} sub-sampling). Moreover, since these generators operate on local patches, they are unable to consider global self-similarity statistics during test time.

In contrast, our method is trained on the single input point cloud with self-supervision, and does not require modeling the specific acquisition / sampling process. Finally, our network is \textit{global}, since it generates a subset of the entire shape rather than patches, which encourages the consolidated features to be consistent with the characteristics of the entire shape.

The idea of using deep neural network as a prior has been studied in image processing~\cite{ulyanov2018deep, DoubleDIP}. These works show that applying CNNs directly on a low resolution or corrupt input image, can in fact learn to amend the imperfect input. 
Similar approaches have been developed for 3D surfaces~\cite{williams2019deep,Hanocka2020p2m}, as a means to reconstruct a consolidated surface. Deep Geometric Prior (DGP) is applied on local charts and lacks global self-similarity abilities. Point2Mesh~\cite{Hanocka2020p2m} builds upon MeshCNN~\cite{Hanocka2019MeshCNN} to reconstruct a watertight mesh containing global self-similarity through weight-sharing. 

Our work shares similarities with a series of recent works that learn from a single input image 
~\cite{ulyanov2018deep,shocher2018zero, zhou2018non, shaham2019singan}.
These works down-sample the input image or local patches and train a CNN to generate and restore the original image or patch, from the down-sampled version. Then the trained network is applied on the original data to produce a higher resolution or expanded output image containing the same features and patterns as the input.
The rationale in these works is similar to ours. A \rg{network} is trained to predict the high feature details of a single input from sub-sampled versions, learning the internal statistics to map low detailed features to high detailed ones. In our work, we partition the input data, rather than simply down-sampling it, to re-balance the desired consolidation criterion in each source and target subset.

\begin{figure}[ht!]
    \centering
        \newcommand{\pl}{-3}
    
      \begin{overpic}[width=\columnwidth]{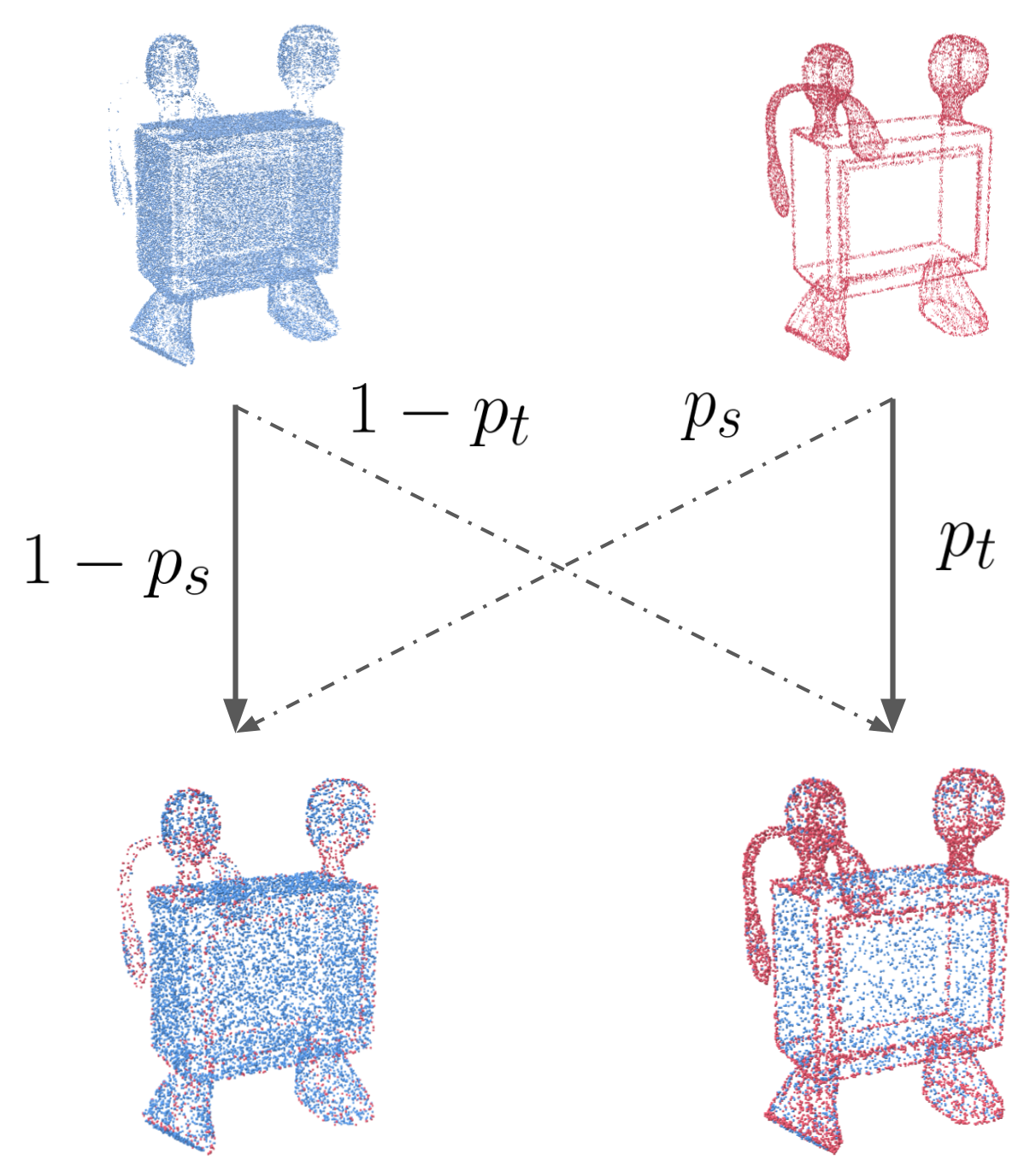}
    \put(18,  \pl){\textcolor{black}{Source}}
    \put(73, \pl){\textcolor{black}{Target}}
    \end{overpic}
    
    \caption{\rh{Approximate (sharp) consolidation illustration. Each point in the input point cloud is categorized as being either \emph{positive} (red) or \emph{negative} (blue) based on a loose approximation of the desired consolidation criterion. The source and target subsets are \textit{re-balanced}, such that the target subset contains mostly positive points (\emph{e.g.,} roughly sharp), enabling the network to learn the desired consolidation characteristic.}}
    \label{fig:sampling_figure}
\end{figure}

\section{Consolidated Point Generation}
\begin{figure}[b]
    \centering
    \newcommand{\pl}{-3}
    \begin{overpic}[width=\columnwidth]{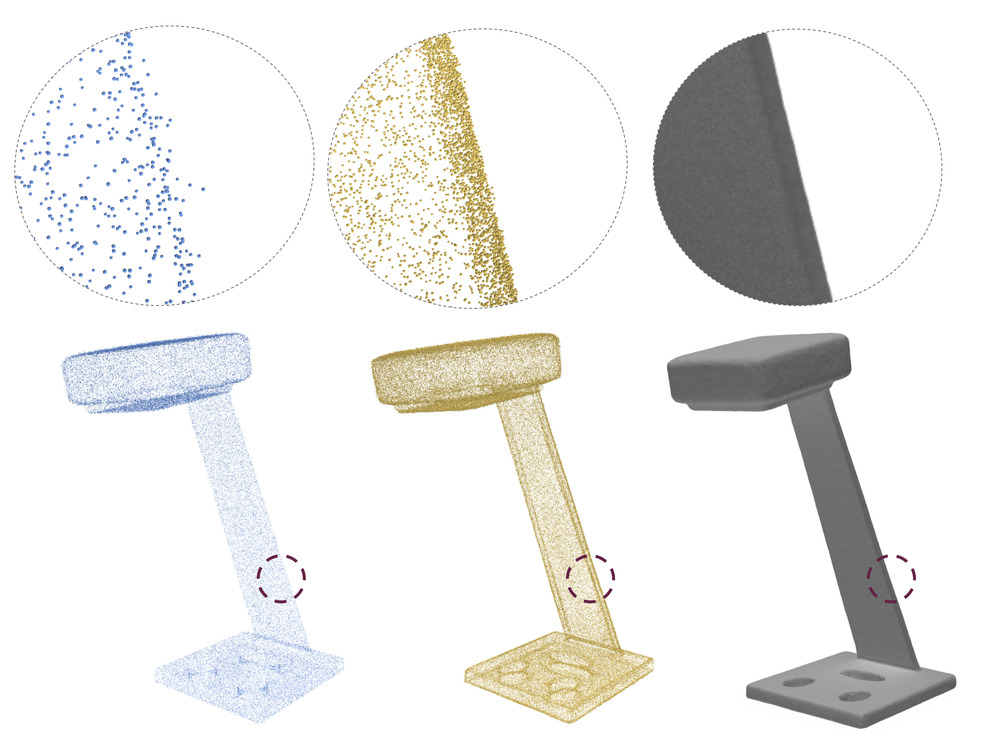}
    \put(17,  \pl){\textcolor{black}{Input}}
    \put(45, \pl){\textcolor{black}{\ourmethod{}}}
    \put(74, \pl){\textcolor{black}{Reconstructed}}
    \end{overpic}
    \caption{Given a noisy input point cloud with outliers, \ourmethod{} learns to consolidate and strengthen sharp features, \rh{which can be used to reconstruct a mesh surface with sharp features.}}
    \label{fig:lamp22}
\end{figure}
We train a network to generate a consolidated set of points using the self-supervision present within a single input point cloud. 
We repeatedly \textit{self-sample} the input point cloud with global subsets which are used to train a deep neural network. We define source and target subsets according to the desired consolidation criterion (e.g., generating sharp points or points in sparse regions). Our training procedure is illustrated in Figure~\ref{fig:overview}.
\begin{figure*}[h]
    \centering
    \newcommand{\pl}{-1}
    \begin{overpic}[width=\textwidth]{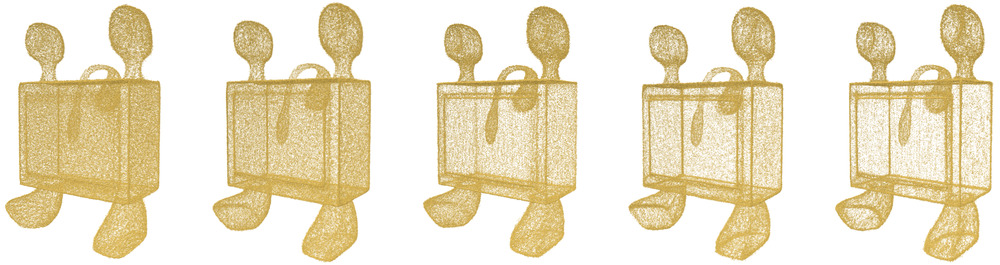}
    \put(6,  \pl){\textcolor{black}{50\%}}
    \put(26, \pl){\textcolor{black}{60\%}}
    \put(48, \pl){\textcolor{black}{70\%}}
    \put(67, \pl){\textcolor{black}{80\%}}
    \put(90, \pl){\textcolor{black}{85\%}}
    \end{overpic}
    \caption{\rh{Results of training on different percentages of \emph{positive} points ($p_{T}$) in the sampled target subsets for sharp feature consolidation. Increasing the amount of positive points in the target subsets results in a sharper predicted output (\emph{e.g.,} 85\%), where as less positive points (\emph{e.g.,} 50\%) results in a more uniform output.}}
    \label{fig:sharpness_prob}
\end{figure*}

The network is trained on disjoint pairs of source and target subsets, and implicitly learns to consolidate the point cloud by generating offsets which are added to the input source points such that they resemble a target subset (\emph{i.e.,} predicted target subset). 
Since the network regresses displacements that are relative to the source subset, the source and target subsets should be disjoint to avoid favoring the trivial solution (predicting zero offsets). 
We use the bidirectional Chamfer distance between the predicted target subset and a target subset to train the network. During inference, the network is fed with random subsets of points from the input, which it displaces to synthesize a consolidated point set (Figure~\ref{fig:inference}). This process is repeated to obtain an arbitrarily large set of consolidated points with an emphasis on the consolidation criterion (\emph{e.g.,} sharp feature points or points in sparsely sampled regions).

We categorize each point in the input via a loose approximation of the desired consolidation criterion, as either \emph{positive} (\emph{e.g.,} sharp), or \emph{negative} group. 
\rg{We focus on two important types of consolidation: generating points on sharp features and inside sparsely sampled regions.
\rev{If the consolidation criterion is \textit{sharp points}, we aim to generate points near edges and sharp regions by balancing the sub sampling according to the local sharpness. If the consolidation criterion is \textit{sparse points}, we aim to generate points in sparse regions by balancing the sub sampling according to the local density.}
We consider a point to be \emph{roughly} sharp if it has a high curvature, or \emph{roughly} low-density if the closest points are far away.}
We use this labeling to repeatedly generate different target subsets with a large number of points from the \emph{positive} group, whereas source subsets contain mostly points from the \emph{negative} group (see Figure~\ref{fig:sampling_figure}). 
\rg{As we show in Section~\ref{sec:exp},}
\rh{this process simultaneously eliminates noise and outliers, as a byproduct of the neural consolidation.}

\subsection{Source and target re-balancing}
\begin{figure}[b]
    \centering
    \newcommand{\pl}{-3}
    \begin{overpic}[width=\columnwidth]{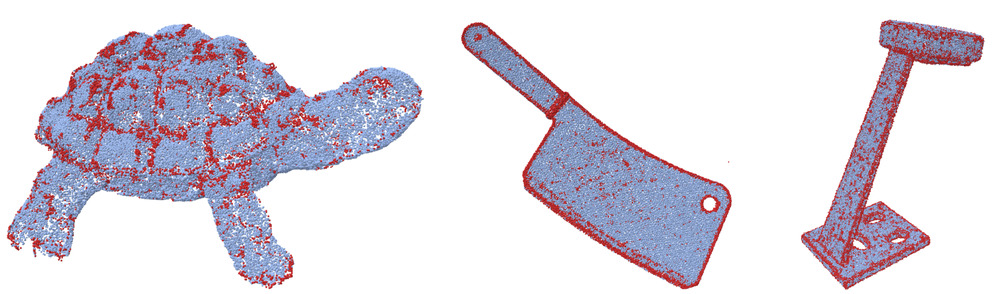}
    \put(15,  \pl){\textcolor{black}{20\%}}
    \put(54, \pl){\textcolor{black}{25.9\%}}
    \put(84, \pl){\textcolor{black}{38.5\%}}
    \end{overpic}
    \caption{\rh{Visualization of the points classified as positive (red) and negative (blue) for sharp feature consolidation. The kNN curvature was thresholded by the mean curvature, resulting in a percentage of positive points which is \emph{sensitive} to the amount of sharp features in the input point cloud.}}
    \label{fig:mean_curve}
\end{figure}
\begin{figure}[hb]
    \centering
    \newcommand{\plt}{49}
    \newcommand{\pl}{-2}
    \begin{overpic}[width=\columnwidth]{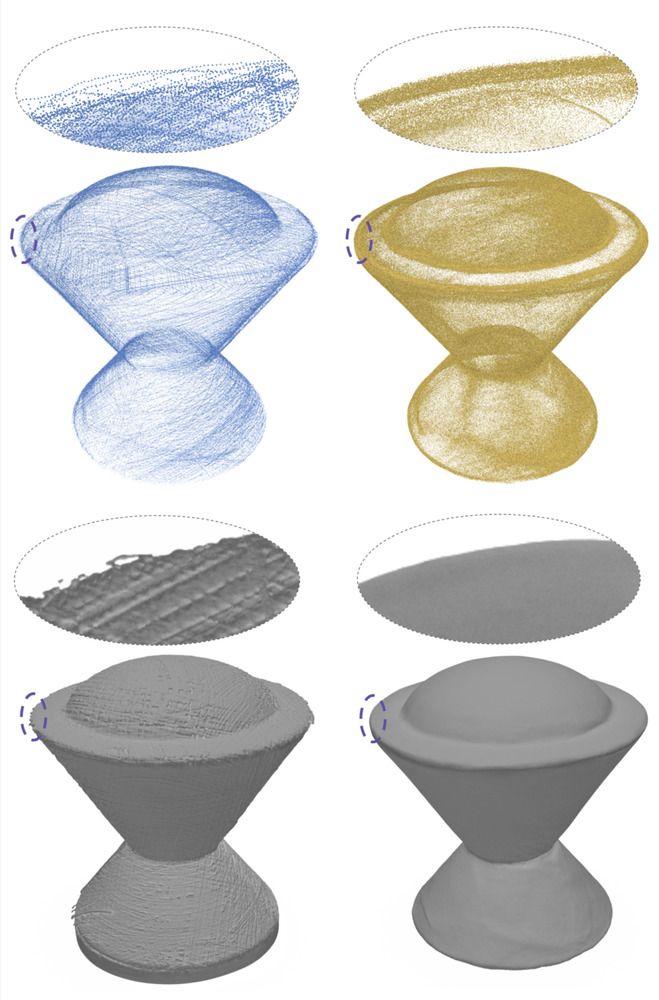}
    \put(16,  \pl){\textcolor{black}{(a)}}
    \put(50, \pl){\textcolor{black}{(b)}}
    \put(15,  \plt){\textcolor{black}{Scan}}
    \put(43, \plt){\textcolor{black}{\ourmethod{}}}
    \end{overpic} \\
    \caption{Scaned input contains artifacts which \ourmethod{} removes while preserving sharp features. Reconstructing a surface \revTwo{using Poisson~\cite{kazhdan2006poisson}} directly from the scanned input (a) yields an undesirable reconstruction, where as reconstructing the surface directly from the output of \ourmethod{} (b) leads to favorable results.}
    \label{fig:door_knob}
\end{figure}

\rh{
We repeatedly sample points from the input point cloud $X$ to create different source $\mathbb{S}$ and target $\mathbb{T}$ subsets. We aim for the target to contain mostly positive points from the input point cloud, based on the desired consolidation criterion. On the other hand, the source subset should contain mostly negative points. However, since the exact categorization is unknown \emph{a priori}, we define proxies for sharpness and sparseness via an estimation of the local curvature and density, respectively. Note, that while \rg{there is a variety of possible} %
methods for approximating the desired consolidation criterion, we observed that our architecture \rg{is} capable of producing favorable consolidation regardless of which method \rg{is} used. %
\rg{Using an} approximate definition of the desired consolidation criterion is more than sufficient, since networks are good at understanding the core modes of the data and ignoring outliers.

These \rg{approximations} %
are used to label each point as either \emph{positive} or \textit{negative}. We re-balance the sampling probabilities of the source and target subsets using the positive and negative point labels. Specifically, each time we sample a target subset, we randomly select $p_T$ percent (\emph{e.g.,} $p_T=85\%$) from the positive set, and the remaining percentage $1 - p_T$ of points are randomly selected from the negative set. Conversely, each source subset is made up of $p_S$ percent of randomly selected negative points, and  $1 - p_S$ percent \rg{of} positive ones (see Figure~\ref{fig:sampling_figure}).}

\subsubsection{Sharpness Approximation}
\rh{Curvature is a simple yet effective heuristic for sharp features~\cite{smith2019geometrics}. We estimate the local curvature of each point, which is used as a proxy for \emph{sharpness}. First, we estimate the normal \rg{of} each point using a simple plane fitting technique. We calculate the difference in normal angles for each point and its k-nearest neighbors ($k$NN). The estimated curvature of a point is given by summing all angle differences with each of its $k$ neighbors. 
}

\subsubsection{Density Approximation}
\rh{We approximate the sampling density of a point based on the distances in each points $k$NN, which is used as a proxy to determine if points are inside \emph{sparsely} sampled regions. For every point $x_i$ we calculate the distance to the $k$NN, and the local sampling density is taken to be  $\frac{k}{{\max \left[  \mathrm{kNN}(x_i)\right] }^3}$.}

\rh{
\subsubsection{Point labeling}
The points are labeled as either \emph{positive} or \textit{negative} according to the desired consolidation criterion (\emph{i.e.,} sharp or sparse).
For the sharp consolidation, we labeled points as sharp (\emph{positive}) if the local curvature of a point was greater than the mean curvature of all points in the input point cloud. On the other hand, points are labeled as non-sharp (\emph{negative}) if the local curvature is less than the mean curvature. Figure~\ref{fig:mean_curve} shows the positive and negative labeling for sharp points for a few shapes.

For the sampling consolidation, we labeled points as being inside sparsely sampled regions (\emph{positive}) using $k$-means. We cluster the estimated density values from the input point cloud using $k$-means, where $k=3$. The points associated with the cluster center with the smallest value are assigned a positive label, and the remaining points are labeled as negative. It should be noted that once the desired consolidation property is estimated (\textit{i.e.,} sharpness or density), the labeling can be performed using either a simple threshold or $k$-means, regardless of the property. While more sophisticated labeling \rg{strategies} can be incorporated, these simple techniques demonstrate that our method does not require perfect labeling to succeed.

}

\subsubsection{Set Selection}
Each point and its associated label are used to create the training source and target subsets $\mathbb{S}, \mathbb{T} \in \mathbb{R}^{m \times 3}$ from the input point cloud  $X \in \mathbb{R}^{n \times 3}$ (where $m < n$). The probability of sampling a \rh{positive} point in the source and target subset is given by a binary (Bernoulli) distribution with probabilities $p_{S}$ and $p_{T}$, respectively. Naturally, we set the probability for drawing \rh{positive} points in the target to be high, whereas the probability for \rh{positive} points in the source is low, $0 \leq p_{S} \leq p_{T} \leq 1$. \rh{Finally, we sample each point from the positive class with uniform probability.} Each time we create a new source and target subset pair, we ensure that they are disjoint $\mathbb{S} \cap \mathbb{T}=\emptyset$, to avoid the trivial solution. The effect of training on different $p_{T}$ is shown in Figure~\ref{fig:sharpness_prob}.

\subsection{Network structure}
The \rh{consolidation network} \rg{$G$} is a fully-convolutional point network. In this work, we use PointNet++\cite{qi2017pointnet++}, which provides the fundamental neural network operators for irregular point cloud data.

The %
\rh{network} $G$ receives a source subset $\mathbb{S} \in \mathbb{R}^{m \times 3}$ and outputs a displacement vector $\Delta \in \mathbb{R}^{m \times 3}$ which is added to $S$, resulting in the predicted target ${\hat{\mathbb{T}} = \mathbb{S} + \Delta}$. The \rh{weights are} trained using a \emph{reconstruction} loss, which compares the predicted target  $\hat{\mathbb{T}}$ to a target subset $\mathbb{T} \in \mathbb{R}^{m \times 3}$.

Each point in the source has a 3-dimensional $<x,y,z>$ coordinate associated with it, which is lifted by the \rh{network} to a higher dimensional feature space through a series of convolutions, non-linear activation units and normalization layers. Each deep feature per point is eventually mapped to the final 3-dimensional vector, which displaces the corresponding input source point to a point which lies on the manifold of target shape features. We initialize the last convolutional layer to output near-zero displacement vectors, which provides a better initialization and therefore more efficient learning.

\subsection{Reconstruction Loss}
The reconstruction loss is given by the bi-directional Chamfer distance between the target $\mathbb{T}$ and predicted target $\mathbb{\hat{T}}$ subsets, which is differential and robust to outliers~\cite{fan2017point}. The \rg{bi-directional} Chamfer distance between two point sets $P$ and $Q$ is given by:
\begin{equation}
\label{eq:chamfer}
    d(P, Q) = \sum_{p \in P} \min_{q \in Q} || p - q ||^{2}_2 + 
    \sum_{q \in Q} \min_{p \in P} || p - q  ||^{2}_2.
\end{equation}
\rg{This} distance is invariant to point \emph{ordering}, which avoids the need to define correspondence. Instead, the network \textit{independently} learns to map source points to the underlying target shape distribution. Since there is no correspondence, it is important that each pair of source and target subsets $\mathbb{S}, \mathbb{T}$ are disjoint, which helps avoid the trivial solution of no-displacement (\emph{i.e.,} $\Delta = \vec{\mathbf{0}}$).

\begin{figure}[b]
    \centering
    \centering
    \newcommand{\pl}{-2}
    \begin{overpic}[width=\columnwidth]{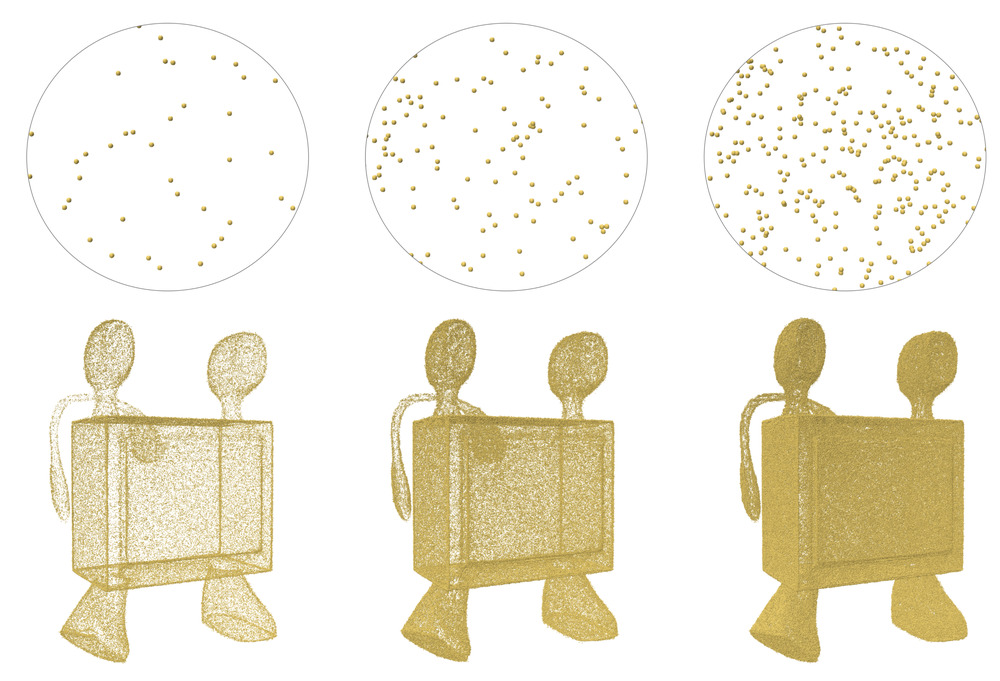}
    \put(12,  \pl){\textcolor{black}{160K}}
    \put(48, \pl){\textcolor{black}{480K}}
    \put(82, \pl){\textcolor{black}{1280K}}
    \end{overpic}
    \caption{Generating a consolidated point cloud at arbitrary resolutions. Increasing the number of points creates greater coverage on the shape surface, rather than simply generating the same points on top of each other.}
    \label{fig:output_resolution}
\end{figure}

\begin{figure}[b]
    \centering
    \newcommand{\pl}{-3}
    \begin{overpic}[width=\columnwidth]{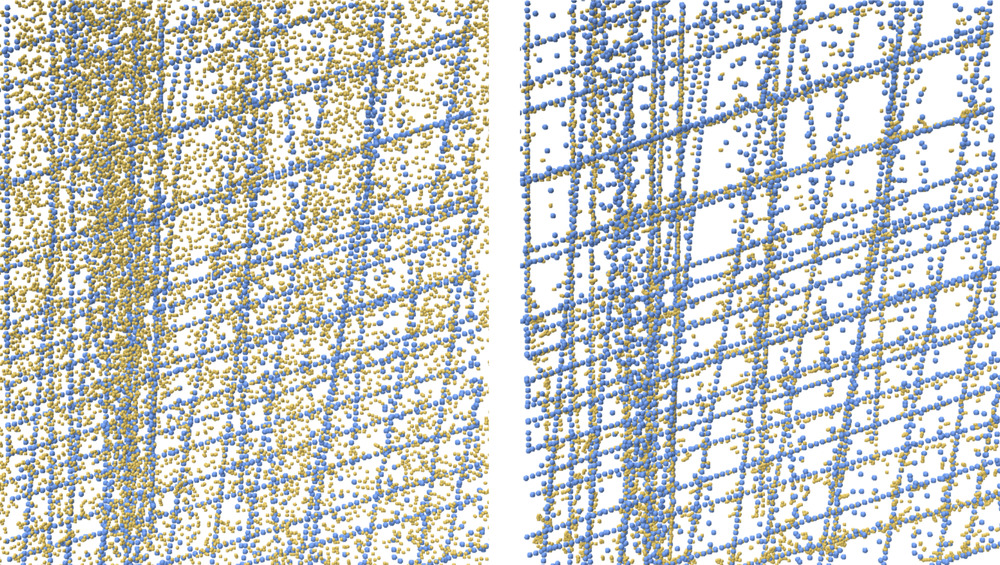} 
        \put(12,  \pl){\textcolor{black}{Shared Weights}}
        \put(65, \pl){\textcolor{black}{Local Network}}
    \end{overpic}
    \caption{Training a local network on a patch of the input (right) compared with training the network from subsets across the input point cloud. Observe that when training on a single patch the network overfits to the data. On the other hand, when trained on the entire shape it generalizes well by sampling the underlying surface.}
    \label{fig:local_network}
\end{figure}

\subsection{Inference}
After training is complete, the \rh{network} has successfully learned the underlying distribution of the target point features. During inference, the \rh{network} displaces subsets from the input point cloud to create novel points which are part of the distribution of the target (\emph{i.e.,} sharp or sparse) features. This process is repeated to achieve an infinitely large amount of samples on the underlying surface. Due to the large network capacity and rich feature representation, novel points can be generated as a function of their input subset. For a given point in the input point cloud $x \in X$, we can generate a \emph{large} variety of target feature points based on the variety of points in the input subset. Therefore, by combining the results of $k$ different subsets of size $m$, we obtain an up-sampled and consolidated version of the input with a size of $k \cdot m$, \rh{with an emphasis on the desired consolidation criterion.
This process is illustrated in Figure~\ref{fig:inference}, where the input subset $S_i$ is passed through $G$ resulting in $T_i$ and $0\leq i \leq k$. 
}

\rh{\subsubsection*{Diversity of Consolidated Points.}}
A key feature of the proposed approach is that the consolidated point cloud can be generated at an arbitrarily large resolution. More specifically, we randomly sample a set of input points from the point cloud, $<x_i, x_j, x_k, \ldots > \; \in X$, which generate offsets $<\Delta_i, \Delta_j, \Delta_k, \ldots >$ that are added back to the input points resulting in the novel samples $<x_i + \Delta_i, x_j + \Delta_j, x_k + \Delta_k, \ldots >$. For a particular point $x_i$, the corresponding offset $\Delta_i$ generated by the network $G$ is a function of the input samples, i.e., $\Delta_i = G(<x_i, x_j, x_k, \ldots >)$. In other words, we can generate a variety of different outputs, based on sampling different input subsets since $\Delta_i = G(<x_i, x_j, x_k, \ldots >) \neq G(<x_i, x_l, x_m, \ldots >)$. Figure~\ref{fig:patches} demonstrates the diversity of such an output set of points (yellow) for a given input point (red) with different subsets as input.
Notably, increasing the number of output points generates greater coverage on the underlying surface, without generating redundant points. To see this further, observe how increasing the resolution in Figure~\ref{fig:output_resolution} results in a denser sampling. Indeed, this is possible since each time a different subset is sampled from the input point cloud, a different set of output points is generated.

\begin{figure}[b]
    \centering
    \newcommand{\pl}{3}
    \begin{overpic}[width=\columnwidth]{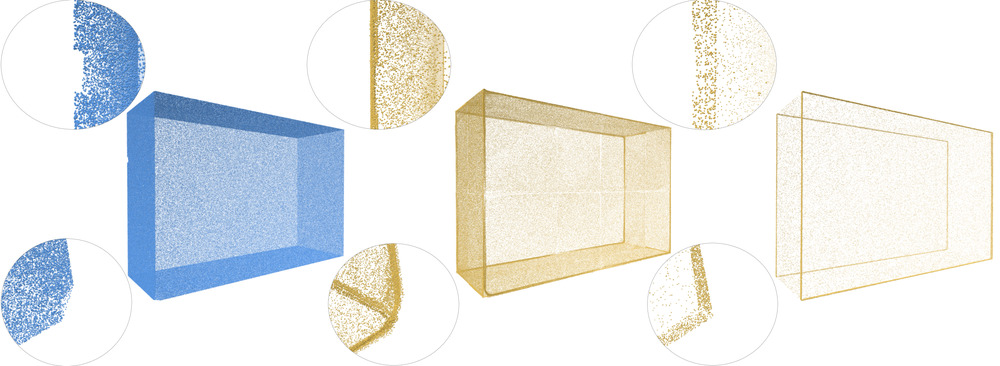}
    \put(22,  \pl){\textcolor{black}{Input}}
    \put(52, \pl){\textcolor{black}{Local}}
    \put(84, \pl){\textcolor{black}{Global}}
    \end{overpic}
    \caption{The power of global shared weights. Training $8\times$ different local networks on different local regions of the input results in overfitting to the missing regions. On the other hand, when using training a single global network on subsets of the input, the shared weights learn important non-local information for completing missing regions.}
    \label{fig:cube8}
\end{figure}

\section{Experiments}
\label{sec:exp}
We conduct a series of qualitative and quantitative experiments to validate the ability of our technique to perform point cloud consolidation and up-sampling. We demonstrate results on a variety of different shapes, which may contain noise, outliers and non-uniform point density on synthetic and real scanned data. We compare against the following methods: EC-Net~\cite{yu2018ec}, PointCleanNet (PCN)~\cite{rakotosaona2019pointcleannet}, EAR~\cite{huang2013edge} and WLOP~\cite{huang2009consolidation}. 

\begin{figure}[h]
    \centering
    \newcommand{\pl}{-4}
    
    \includegraphics[width=0.8\columnwidth]{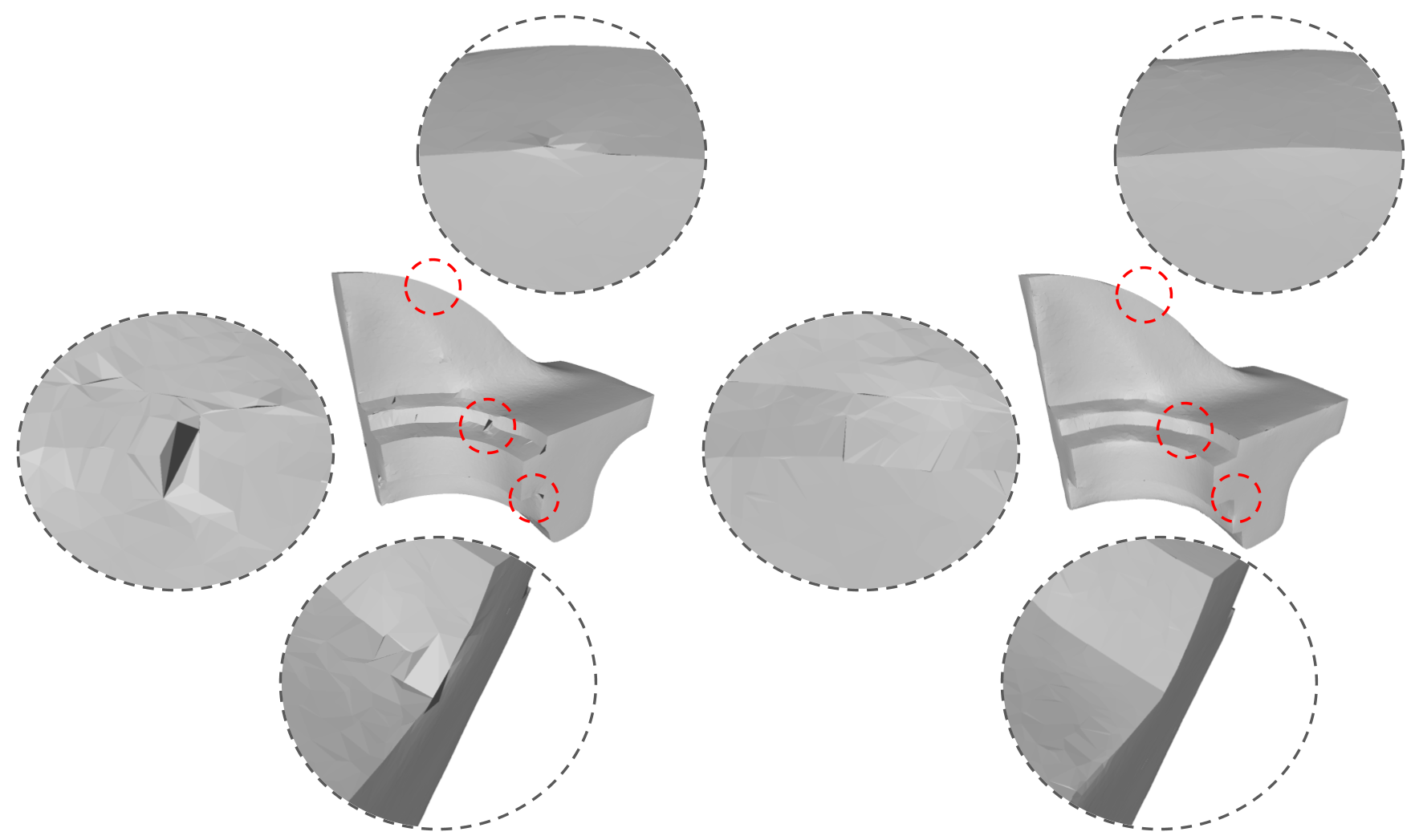}
    \includegraphics[width=\columnwidth, trim=0 0 0 0 ]{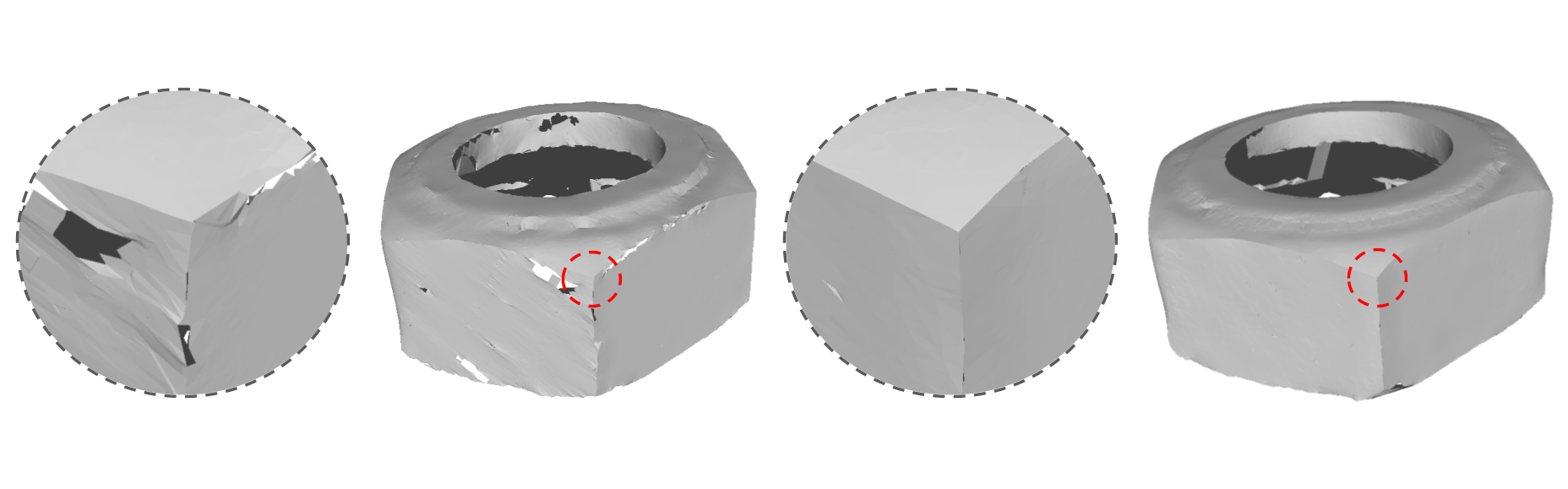}
\includegraphics[width=\columnwidth, trim=0 0 0 0 ]{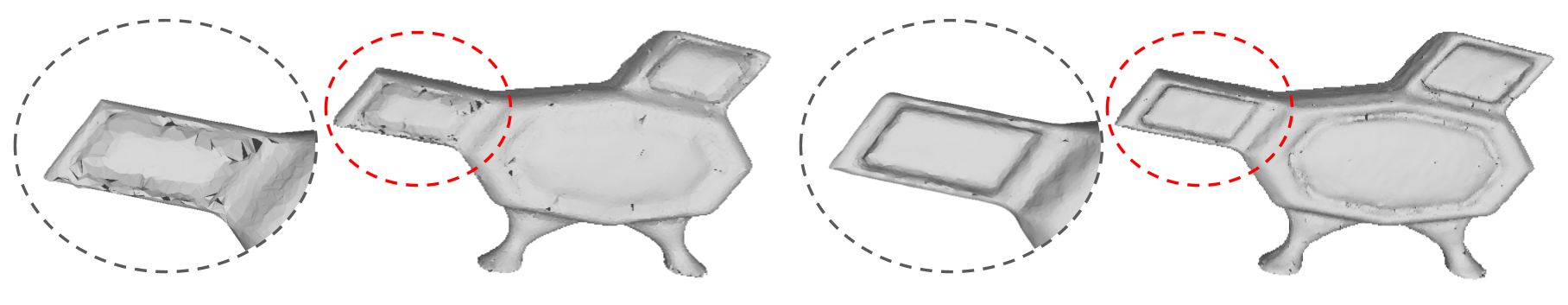}
    \begin{overpic}[width=\columnwidth, trim=-12cm 0 0 0 ]{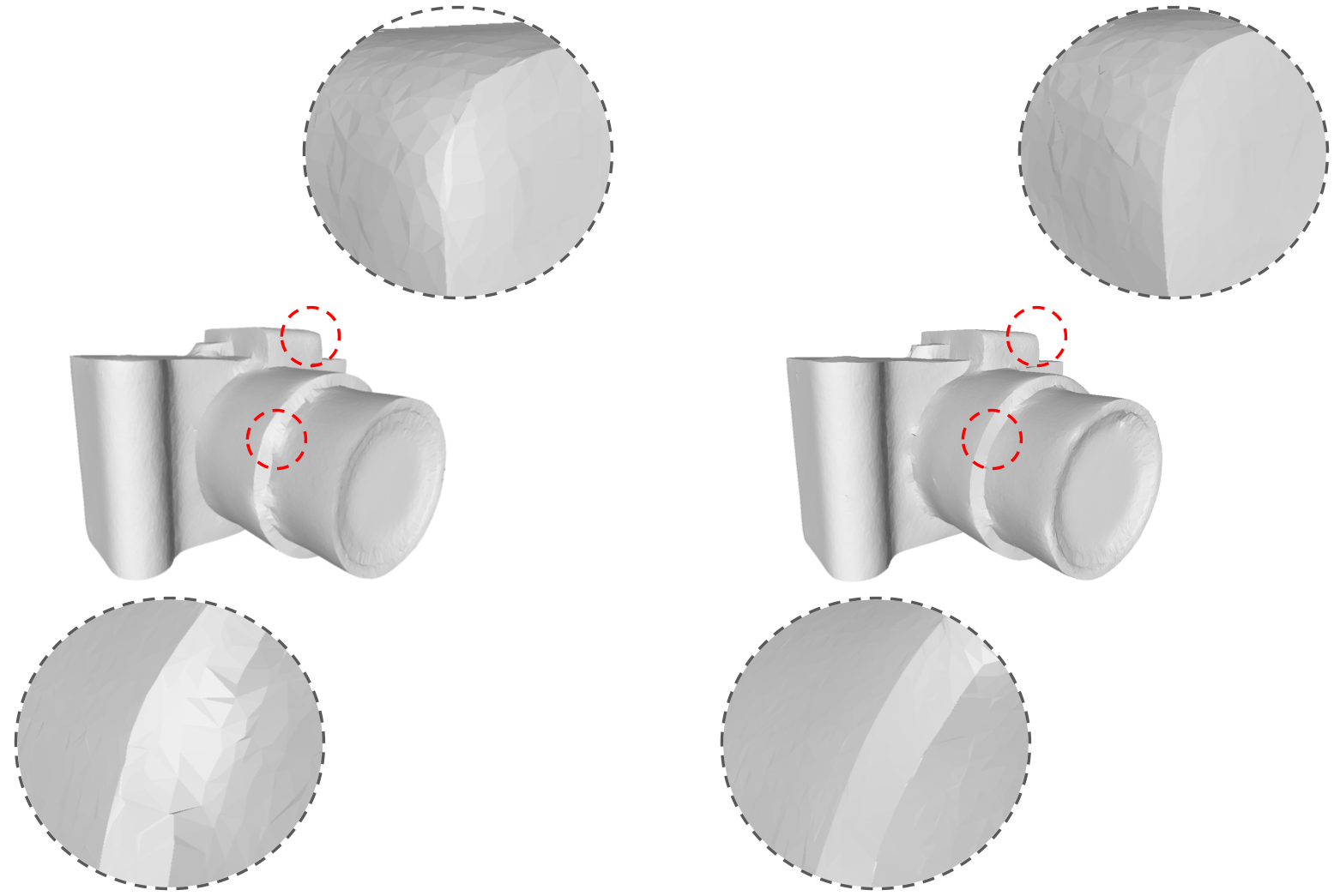}
    \put(30,  \pl){\textcolor{black}{Input}}
    \put(70, \pl){\textcolor{black}{Ours}}
    \end{overpic}
    \caption{\rh{Reconstruction from the input point cloud (left),
    and reconstruction on the output of our method. Observe how consolidation removes noise and preserves sharp features. Mesh reconstruction was performed using~\cite{bernardini1999ball}, and ~\cite{sun2007fast}.}}
    \label{fig:mesh_recon}
\end{figure}

The experiments focus on illustrating the following aspects: global weight-sharing, sharp feature consolidation \rev{(Figure~\ref{fig:mesh_recon})} and up-sampling, sparse region consolidation and unification and denoising. Unless otherwise stated, all results of our approach in this paper are the direct output from the network, without any pre-processing to the input point cloud or output point cloud. 

\subsection{Weight Sharing} 
We claim that neural networks with weight-sharing abilities posses an inductive bias, enabling removing noise and outliers and complete low density regions. This is in contrast to multiple \textit{local} networks which operate on small patches, without any global context or shared information among different patches. To validate this claim, we train a single global network (which shares the same weights for all subsets) compared against multiple local networks each trained on separate local regions of the input point cloud. The input is a synthetic cube which has missing samples near some corners and edges. The global network is trained on global subsets from the entire cube. We train 8 different local networks on a different octant of the cube. In order to provide each local network with as much data as the single global network, we provide $8\times$ as many points to each octant. Observe the results in Figure~\ref{fig:cube8}. Despite using $8\times$ as many points (in total) and $8\times$ as many networks, the local networks easily overfit to the missing regions. In particular, notice the cube corner. On the other hand, the shared weights exploit the redundancy present in a single shape, and encourages a plausible solution. \rev{Another possibility for using \textit{global} shared weights would have been to partition the shape into many different overlapping patches, and then train a single network (\emph{i.e.,} shares weights) across each of the different patches. We believe this direction could be an interesting avenue to explore in future work.}

We perform another comparison on a real scan from AIM@SHAPE-VISIONAIR. In this setup, we train a single local network on an extracted patch from the scan. The result is shown in Figure~\ref{fig:local_network}. As expected, the local network overfits the artificial pattern from the input scan. For reference, we included the result from the global network that was trained on the entire shape.
\begin{figure}
    \newcommand{\pl}{-2}
    \centering
    \includegraphics[width=\columnwidth]{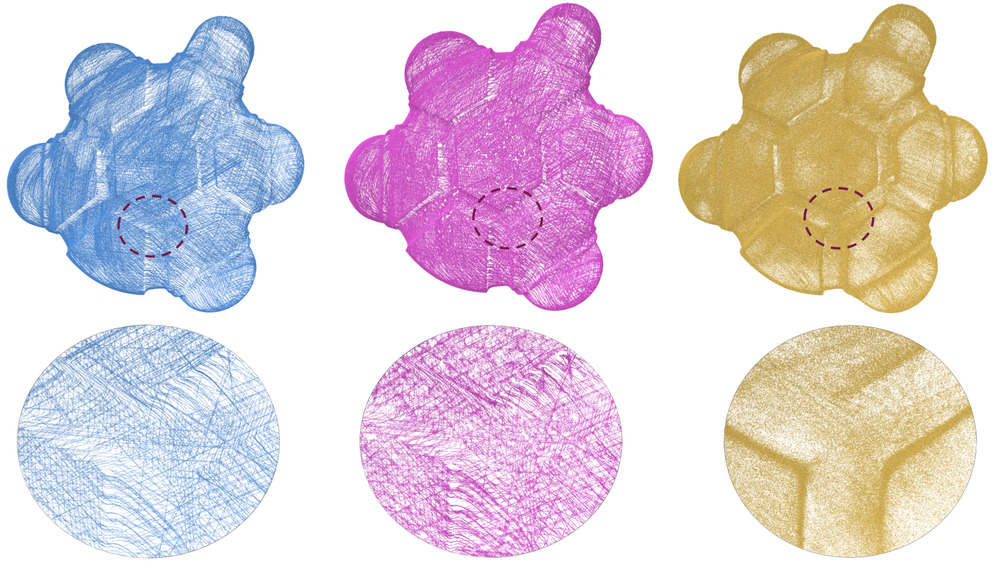} \\
    \includegraphics[width=\columnwidth]{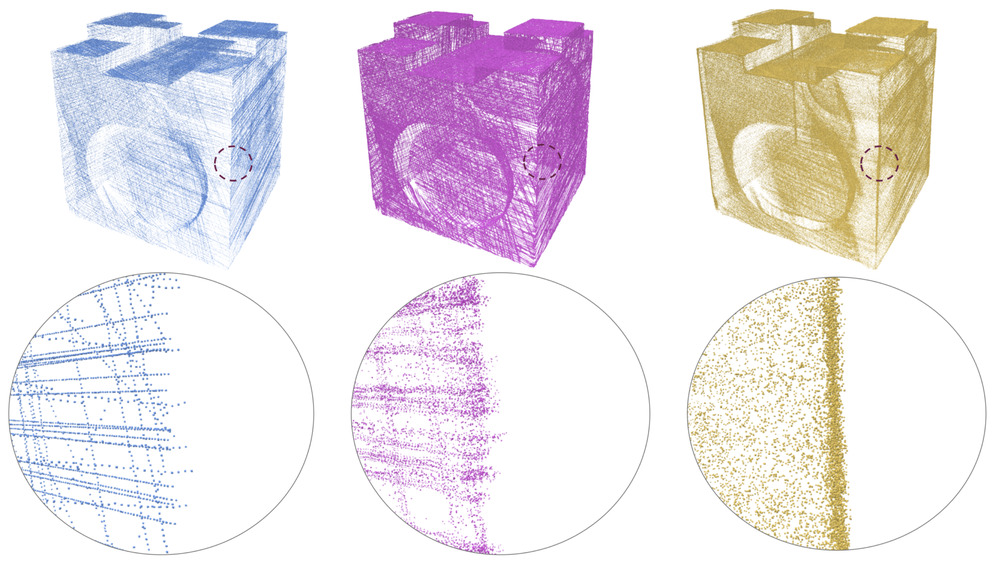} \\
    \begin{overpic}[width=\columnwidth]{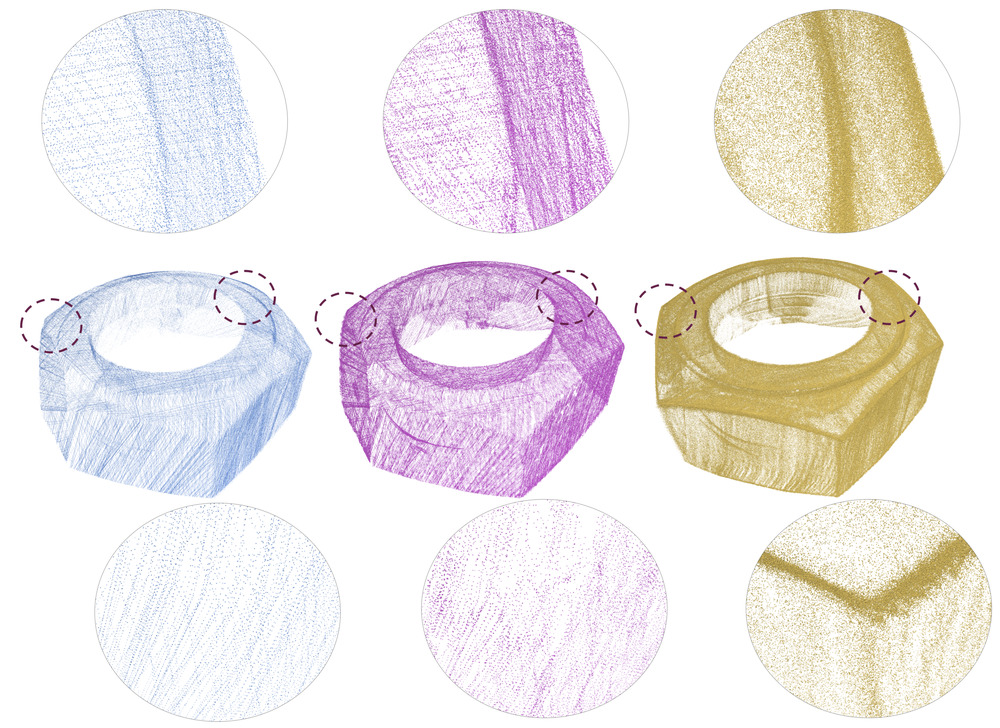}
    \put(10,  \pl){\textcolor{black}{Scan}}
    \put(42, \pl){\textcolor{black}{EC-Net}}
    \put(72, \pl){\textcolor{black}{\ourmethod{}}}
    \end{overpic}
    \caption{Real 3D scans from AIM@SHAPE-VISIONAIR. Observe how \ourmethod{} is agnostic to the unnatural pattern in the input point cloud, by producing a dense consolidated point set with sharp features.}
    \label{fig:real_scans}
\end{figure}

\subsection{Sharp Feature Consolidation}

\subsubsection{Output Point Set Distribution}
\begin{figure}[b]
    \centering
    \includegraphics[width=\columnwidth]{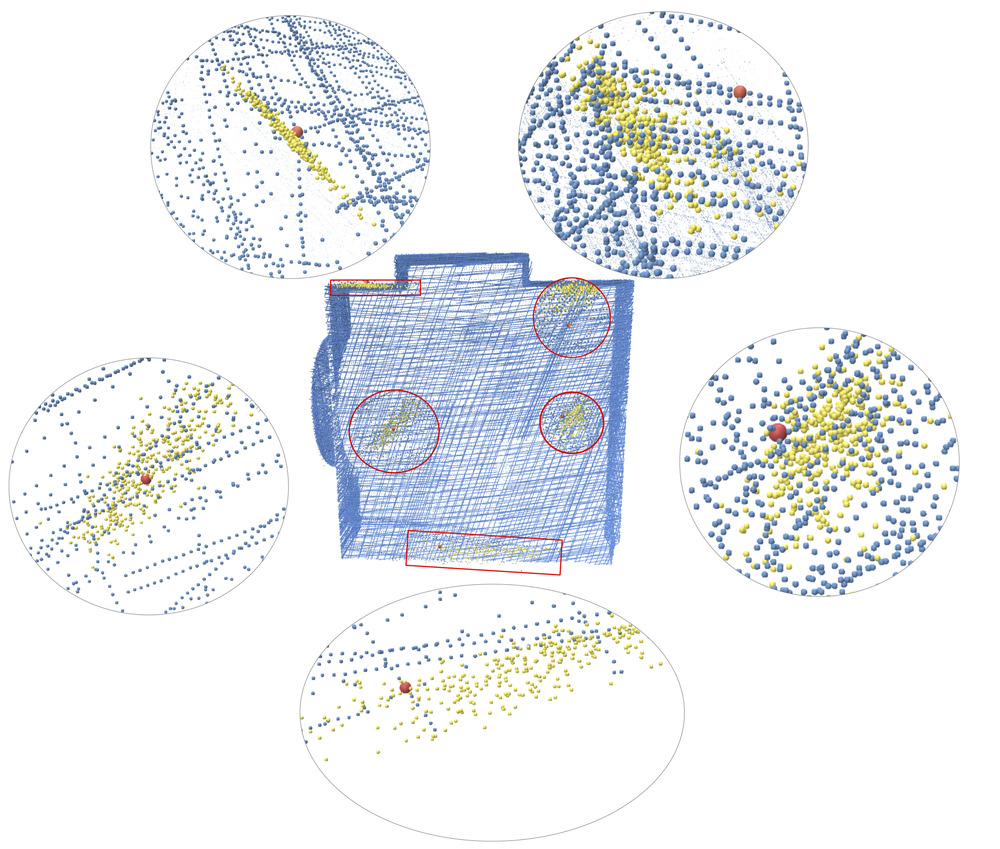} 
    \caption{Diversity in novel sharp feature generation. Given an input point (red) and random subsets from the input point cloud (entire point cloud in blue), our network generates different displacements which are added to the input point to generate a variety of different output points (yellow). Novel points on the surface are \emph{conditioned} on selected input subset.}
    \label{fig:patches}
\end{figure}

Figure~\ref{fig:patches} helps illustrate what the network has learned and the type of diverse outputs it can produce. We selected five (constant) points (red) from the input point cloud (blue), and concatenated them into many random subsets selected during the up-sampling process. Each of the output points (colored in yellow) were produced by different subsets of the input point cloud. 

Observe that the network learns the underlying distribution of the surface during training time, such that during inference an input point can be mapped to many novel output points, which densely sample the underlying surface. Note the top left inset \rev{in Figure~\ref{fig:patches}}, which highlights the network's ability to capture the distribution of an edge, despite the unnatural sampling in the scanned data. Moreover, in the bottom inset, new points are densely generated along the edge boundary of the object, yet, do not surpass the edge itself (remain contained on the surface).

\begin{figure}
    \centering
    \newcommand{\pl}{2}
    \begin{overpic}[width=\columnwidth]{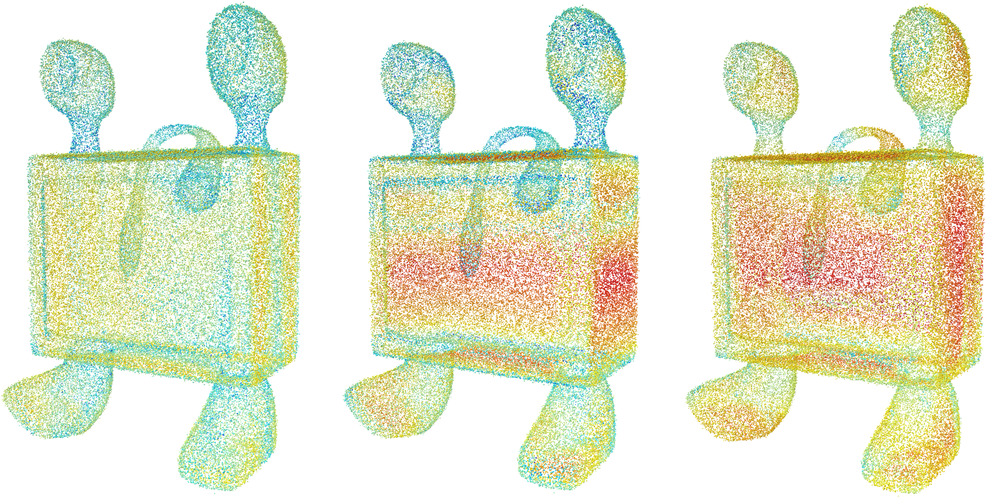}
    \put(10,  \pl){\textcolor{black}{50\%}}
    \put(45, \pl){\textcolor{black}{70\%}}
    \put(79, \pl){\textcolor{black}{85\%}}
    \end{overpic}
    \caption{Heatmap visualization of the network predicted displacements on an input point cloud. We train on different percentages of \emph{sharp} points in the sampled target subsets (results in Figure~\ref{fig:sharpness_prob}). Red and blue are for large and small displacements, respectively. Re-weighting the target sampling probability $p_T$, in order to include more curvy points implicitly encourages the network to predict larger displacements. Observe how the network trained with $p_T=85\%$ displaces the input points inside large flat regions the farthest.}
    \label{fig:displace_heatmap}
\end{figure}

The amount of points generated on sharp features can be controlled via the sampling probability of curvy points in the target subset $p_T$. As shown in Figure~\ref{fig:displace_heatmap}, this directly affects the displacement magnitude of each input point. For example, when $p_T$ is high (\emph{e.g.,} $p_T = 85 \%$), the network generates large displacements for points on flat surfaces.

\subsubsection{Real Scans} 
We test our approach on real scans provided by AIM@SHAPE-VISIONAIR. Figure~\ref{fig:door_knob} demonstrates the ability of our approach to generate sharp and accurate edges corresponding to the underlying shape, despite the presence of unnatural scanner noise. In addition, applying surface reconstruction directly to the input point cloud leads to undesirable results with scanning artifacts and non-sharp edges. On the other hand, reconstructing a surface from the consolidated output of the network leads to pleasing results, with sharp edges. Note that we did not apply any post-processing to our point cloud, however, we did use an off-the-shelf tool in Meshlab to estimate the point cloud normals. 

\rg{Figure~\ref{fig:real_scans} shows} additional results on real scans from AIM@SHAPE-VISIONAIR including a comparison to EC-Net. Observe how our approach is able to learn the underlying surface of the object despite the artificial scanning pattern. Moreover, it produces a dense consolidated point set with sharp features. On the other hand, since EC-Net was not trained on such data, it is sensitive to this particular sampling and generally preserves the artificial scanning pattern in the output. Additional results on real scans can be found in the supplementary material.

\begin{table}
\adjustimage{height=3.7cm,valign=c}{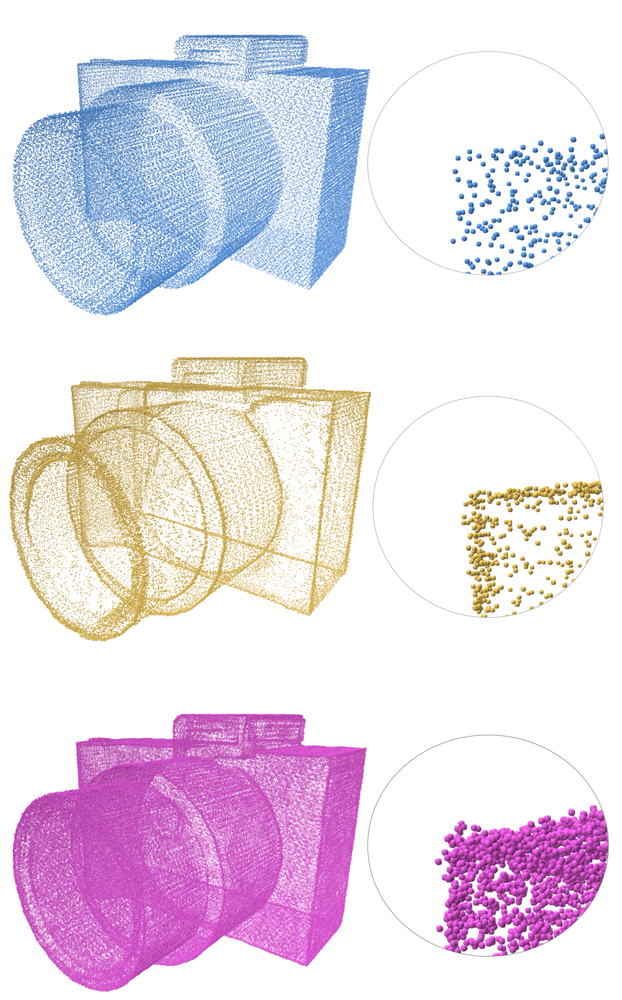}\quad%
     \begin{tabular}{c|c|c|c} %
    \hline
    {\textbf{Shape}} & {\textbf{ours}} & {\textbf{EC-E}} & {\textbf{EC-A}} \\
    \hline\hline %
    camera & \textbf{98.49} & 87.59 & 94.52 \\ 
    fandisk & \textbf{99.86} & 85.97 & 99.02 \\ 
    switch pot & \textbf{99.26} & 90.51 & 89.62 \\ 
    chair & \textbf{99.99} & 97.15 & \textbf{99.99} \\ 
    table & \textbf{99.59} & 97.11 & 95.42 \\ 
    sofa & 68.67 & \textbf{85.95} & 77.92 \\ 
    headphone & \textbf{99.96} & 98.91 & \textbf{99.96} \\ 
    monitor & \textbf{99.92} & 71.17 & 99.7 \\ 
      [1ex] %
    \hline %
    \end{tabular}
\caption{\rh{F-score comparison on the EC-Net dataset. The F-score is w.r.t. to the ground-truth \textit{edge sampled} mesh. Results reported for our method, EC-Net edge points (EC-E), and EC-Net all points (EC-A).}}
\label{table:quant}
\end{table}
\subsubsection{Comparisons} 
\begin{figure*}
    \centering
    \newcommand{\pl}{-2}
    \begin{overpic}[width=16cm]{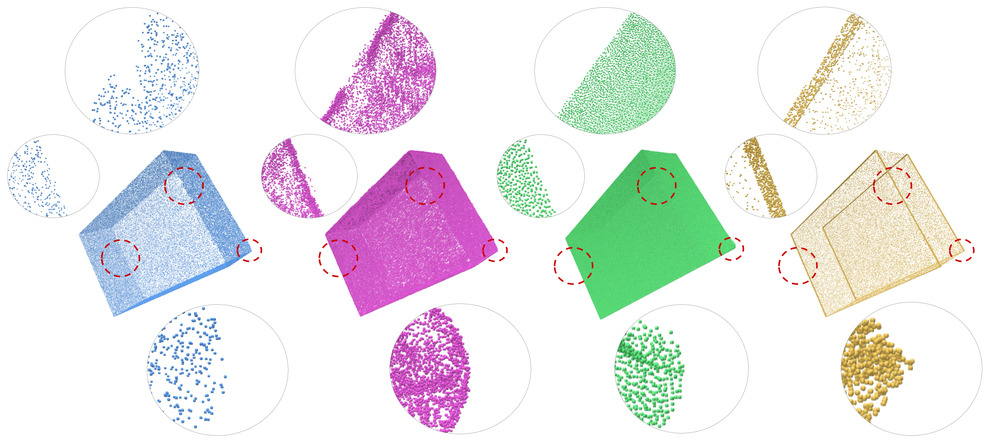}
    \put(12,  \pl){\textcolor{black}{Input}}
    \put(37, \pl){\textcolor{black}{EC-Net}}
    \put(61, \pl){\textcolor{black}{EAR}}
    \put(85, \pl){\textcolor{black}{\ourmethod{}}}
    \end{overpic}
    \caption{Input point cloud with missing samples on sharp features. \rg{The results} shown are the outputs from EC-Net (without edge selection), EAR and our network. Our results are \rg{the} direct result of the network, \dc{without any pre/post processing}.}
    \label{fig:cube_comparison}
\end{figure*}
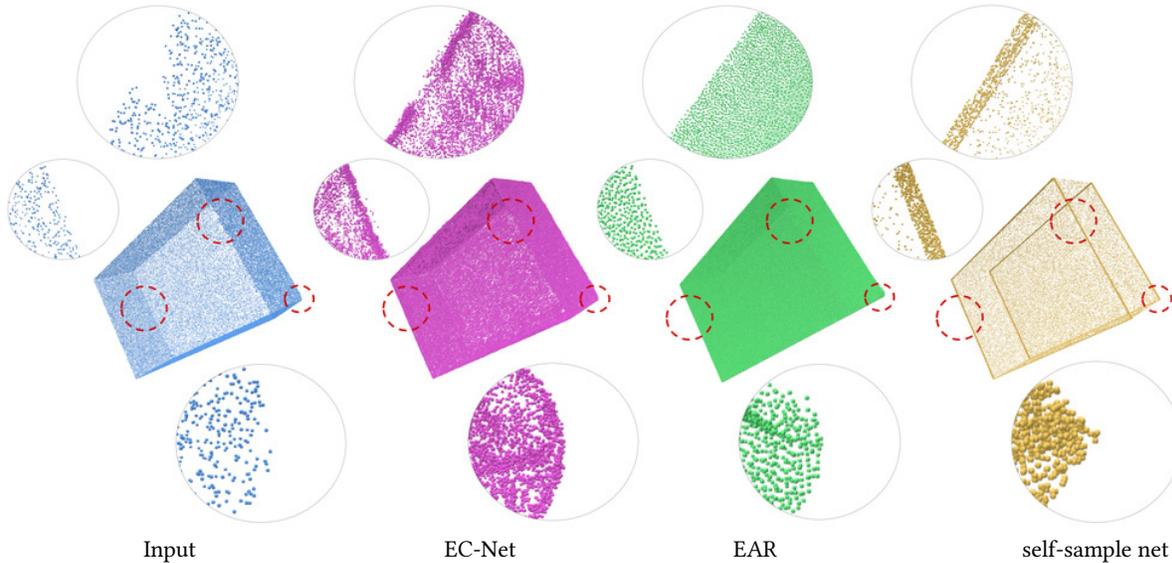

We performed additional qualitative comparisons to EAR and EC-Net.
Figure~\ref{fig:cube_comparison} highlights where our method reconstructs sharp points in a better way compared to these approaches. In particular, the weight-sharing property enables consolidating sharp regions and ignoring noise and outliers. For example, in Figure~\ref{fig:cube_comparison} our approach completes the corner with many sharp points, where the alternative approaches have rounder corners.
\begin{figure}
    \centering
    \newcommand{\pl}{-3}
    \begin{overpic}[width=\columnwidth]{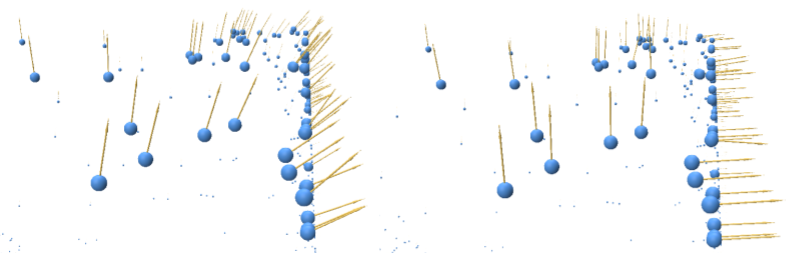}
    \put(20,  \pl){\textcolor{black}{Input}}
    \put(60, \pl){\textcolor{black}{Consolidated}}
    \end{overpic}
    \caption{Normal estimation near edges. Left: a direct PCA-based normal estimation on the input. Right: the same normal estimation on the input points together with the new generated edge points. The new generated points are not present in the figure for a better visual comparison.}
    \label{fig:sharp_normals}
\end{figure}

We perform a quantitative comparisons on the benchmark of EC-Net~\cite{yu2018ec}, which includes eight virtually scanned objects. We use the F-score metric~\cite{Knapitsch2017,Hanocka2020p2m}, which is the harmonic mean between the precision and recall at some threshold $\tau$:
\begin{equation}
F(\tau) = \frac{2P(\tau)R(\tau)}{P(\tau) + R(\tau)}.
\end{equation}
The F-score is computed with respect to an edge-sampled version of the ground-truth the estimated point samples lie on edges ($P(\tau)$), as well as how well all edges have been \textit{covered} ($R(\tau)$).

\rg{
\begin{wrapfigure}{l}{0.05\textwidth}
\centering
\includegraphics[width=0.08\textwidth]{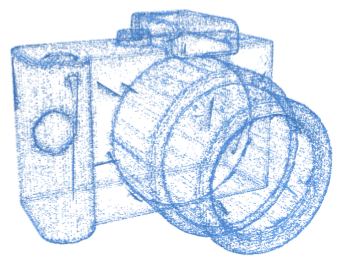}
\centering
\vspace{-10pt}
\end{wrapfigure}
We use an edge-aware sampling method for the ground-truth meshes to emphasize sharp edges in the quantitative comparisons. An edge on the mesh is considered sharp if its dihedral angle is small. We select edges based on their sharpness and length. After an edge is selected, we sample points on the edge adjacent faces with high probability, and with a decreasing probability when moving away from the edge (\emph{i.e.,} a type of triangular Gaussian distribution).

Table~\ref{table:quant} presents the F-score metrics for our approach and two variations of EC-Net provided in the authors reference implementation (edge selected and all output points). On the left side of Table~\ref{table:quant} we show an example from the benchmark: input point cloud (blue), ours (yellow) and EC-Net (all points) in purple. It should be noted that our method does not use any pre or post processing on the input or output point cloud. These qualitative and quantitative comparisons indicate that our method achieves on-par or better results compared to EC-Net despite being trained only on the input shape.}
\begin{figure}[b]
    \includegraphics[width=0.8\columnwidth]{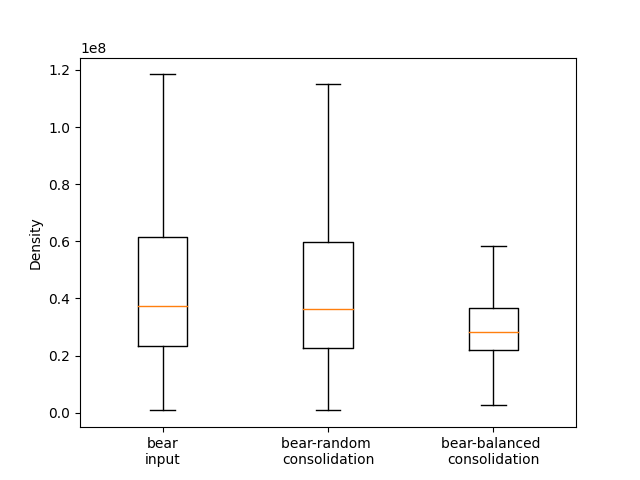}
    \caption{\gm{Boxplot of the per point density estimation. Observe how the balanced consolidation results in a more uniform point sampling across the entire point set (smaller variance)}.}
    \label{fig:box_plot}
\end{figure}

\subsubsection{Edge Points Normal Estimation}
Normals are incredibly
\begin{wraptable}{r}{5.5cm}
\centering
\begin{tabular}{c|c|c|c} %
\hline
{\textbf{Shape}} & {\textbf{input}} & {\textbf{\rev{bilat}}} & {\textbf{ours}} \\
\hline\hline %
fandisk & 23.72 & 8.49 & 4.13 \\
camera & 20.41 & 9.42 & 5.21 \\
chair & 31.03 & 8.8 & 4.59 \\
headphone & 26.56 & 8.66 & 5.03 \\
monitor & 26.46 & 9.18 & 3.96 \\
sofa & 26.35 & 8.34 & 3.22 \\
switch pot & 23.9 & 6.45 & 2.78 \\
table & 25.35 & 8.13 & 3.19 \\
[1ex] %
\hline %
\end{tabular}
\caption{Median normal error in sharp regions (degrees) using PCA normal estimation. \rev{Bilateral filtering is applied on the input in \emph{bilat} and on our consolidated output.} \revTwo{Details about the bilateral filtering are specified in the supplementary.}}

\label{table:sharp_normals}
\vspace{-5pt}
\end{wraptable} 

difficult to estimate, especially in edge regions. Edge regions change non-continuously, therefore, nearby points do not necessarily belong to the same surface on the underlying shape. Estimating normals
according to the nearest neighbors in these situations often results in normals pointing somewhere between the two planes that make up the edge, as can be seen in Figure~\ref{fig:sharp_normals} (input). This defines a new false plane that is not present in the ground truth shape. Our method's ability to generate many new points in the edge region justifies the use of the nearest neighbors normal estimation, because it lowers the probability of points to have neighbors that are over the edge. This results in a much more accurate normal estimation as can be seen in Figures~\ref{fig:sharp_normals}.
Table~\ref{table:sharp_normals} measures the median angle error of normal estimation in edge region. Observe that our generated edge points help decrease the estimation error, sometimes by as much as 75\%.

\subsection{Sparse Region Consolidation}
\begin{figure}[h!]
    \centering
    \newcommand{\pl}{-4}
    
    \includegraphics[width=0.7\columnwidth]{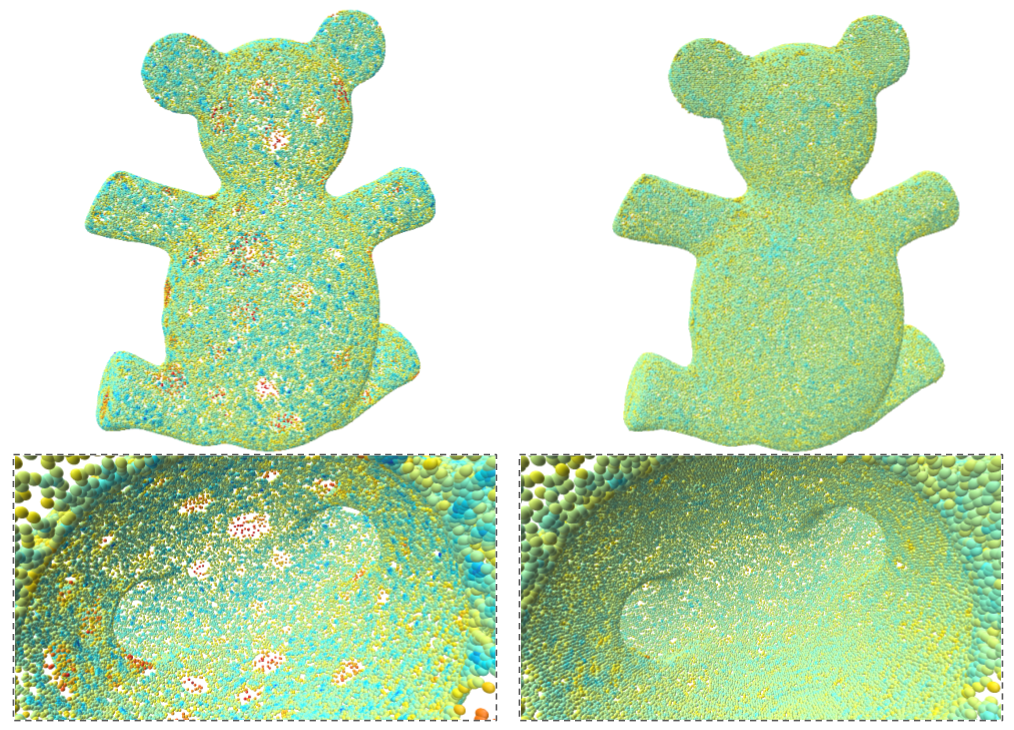}

    \begin{overpic}[width=0.7\columnwidth, trim=0 0 0 0 ]{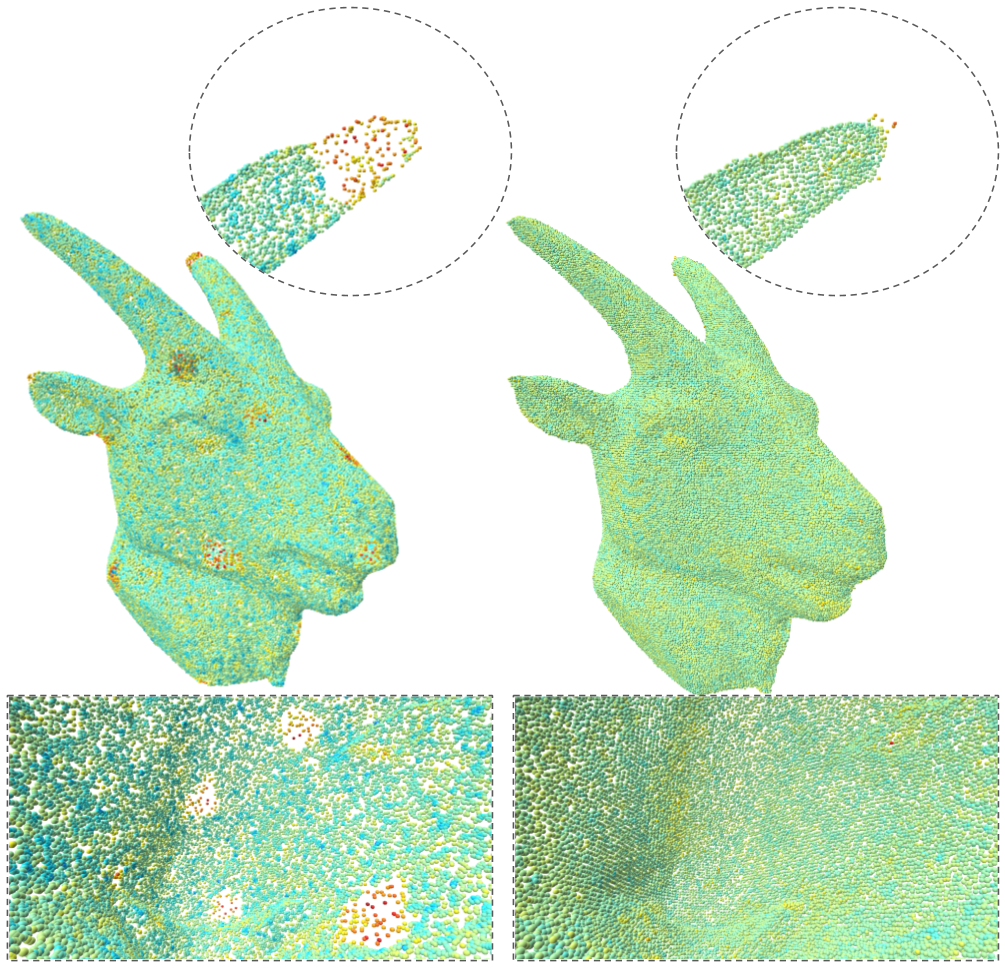}
    \put(20,  \pl){\textcolor{black}{Input}}
    \put(60, \pl){\textcolor{black}{Consolidated}}
    \end{overpic}
    \caption{\gm{Consolidation of sparse regions in point clouds that were unevenly sampled from meshes. The color heat map represents the estimated density, where \textit{hot} corresponds to \textit{sparse} and \textit{cool} corresponds to \textit{dense}.  }}
    \label{fig:density_synt_pc}
\end{figure}

\subsubsection{Density Evaluation Table} 
To obtain a quantitative evaluation of point densification in sparse regions, we ran our method on nine different unevenly sampled point clouds from meshes, some of which can be seen in Figure~\ref{fig:density_synt_pc}.
Each shape was consolidated and up-sampled, and then down-sampled back to the original size for a fair comparison to the input.
Furthermore a box graph of the estimated density value was calculated for each shape, one of which can be seen in Figure~\ref{fig:box_plot}.
Table~\ref{table:density} shows the 25\% to 75\% range of all box graphs that appear in the supplementary material.
To ensure that density re-balancing is crucial for our method, we evaluated the results to ones that were trained with completely random subsets (\textit{i.e.,} not balanced according to sparseness).
As expected, training on density balanced subsets resulted in significant decrease in the range of estimated densities, resulting in more uniformly sampled results. Note that the random subsets did achieve some decrease in range, due to the inherent consolidation properties of the network in upsampling.
In addition we compared against WLOP~\cite{huang2009consolidation}, which failed to move points into the sparse regions and actually worsens the global uniformity of the sampling, as can be seen in Table~\ref{table:density}.

\begin{table}[h]
     \begin{tabular}{c|c|c|c|c} %
    \hline
    {\textbf{Shape}} & {\textbf{Input}} & {\textbf{WLOP}} & {\textbf{Random}} & {\textbf{Balanced}} \\
    \hline\hline %
Bull & 2.17e+08 & 3.13e+08 & 1.07e+08 & \textbf{2.81e+07} \\ 
Ankylosaurus & 8.18e+08 & 3.27e+09 & 4.32e+08 & \textbf{1.03e+08} \\ 
Bear & 3.81e+07 & 4.08e+07 & 3.70e+07 & \textbf{1.45e+07} \\ 
Spot & 2.92e+07 & 3.14e+07 & 2.68e+07 & \textbf{1.16e+07} \\ 
Netta & 4.55e+07 & 4.95e+07 & 3.51e+07 & \textbf{1.14e+07} \\ 
Tiki & 5.63e+07 & 6.05e+07 & 4.38e+07 & \textbf{1.97e+07} \\ 
Candle & 2.02e+08 & 2.69e+08 & 5.66e+07 & \textbf{3.27e+07} \\ 
Goathead & 3.35e+07 & 3.60e+07 & 3.13e+07 & \textbf{2.17e+07} \\ 
Camel & 1.68e+08 & 1.97e+08 & 3.42e+07 & \textbf{2.31e+07} \\
      [1ex] %
    \hline %
    \end{tabular}
\caption{Density 25\% 75\% percentile range. The less the more uniformly distributed, less is better.}
\label{table:density}
\end{table}

\subsubsection{Mesh Reconstruction}
\begin{figure}[h]
    \centering
    \newcommand{\pl}{-4}
    \newcommand{\wdd}{0.8\columnwidth}
    \includegraphics[width=\wdd]{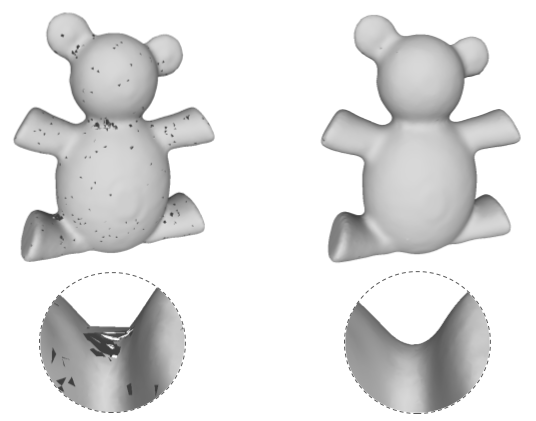}

    \begin{overpic}[width=\wdd, trim=0 0 0 0 ]{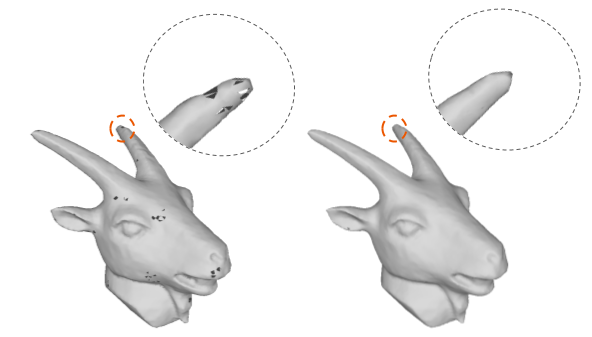}
    \put(15,  \pl){\textcolor{black}{Input}}
    \put(55, \pl){\textcolor{black}{Consolidated}}
    \end{overpic}
    \caption{\gm{Mesh reconstruction results of non-uniformly sampled point clouds (left) and their consolidated version (right). Observe how reconstruction quality improves and holes are filled in the consolidated point set. }}
    \label{fig:density_synt_mesh}
\end{figure}

Figure~\ref{fig:density_synt_mesh} shows mesh reconstruction results using the ball pivot algorithm\rev{~\cite{bernardini1999ball}} of the three consolidated point clouds that appear in Figure~\ref{fig:density_synt_pc}.
Observe that a direct reconstruction of the input results in many holes along the surface, and can also create incorrect surface completions as can be seen in first mesh reconstruction of the bear example.
Our consolidated point sets eliminate the issues caused by the scarcity of points in certain regions, resulting in a higher quality mesh reconstruction. 
\begin{figure}[h]
    \centering
    \newcommand{\pl}{-4}
    
    \includegraphics[width=\columnwidth]{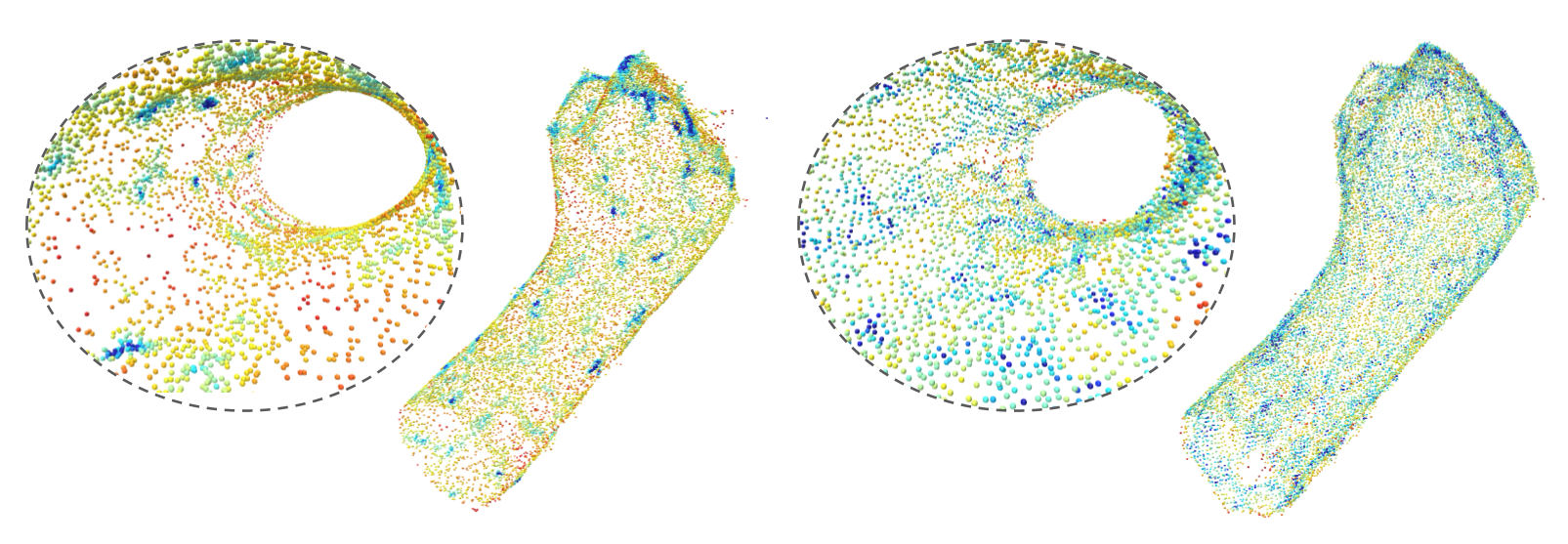}
    \includegraphics[width=\columnwidth, trim=0 0 0 0 ]{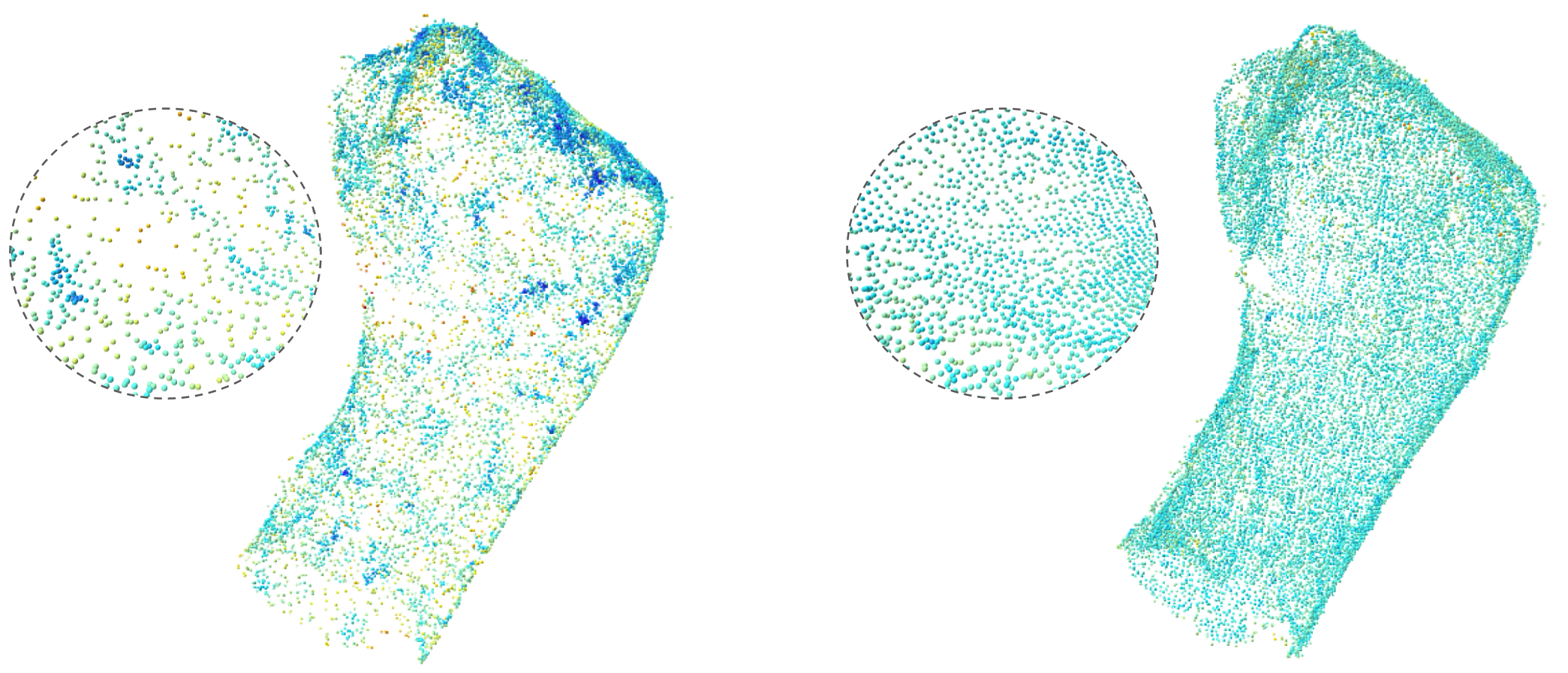}

    \begin{overpic}[width=\columnwidth, trim=0 0 0 0 ]{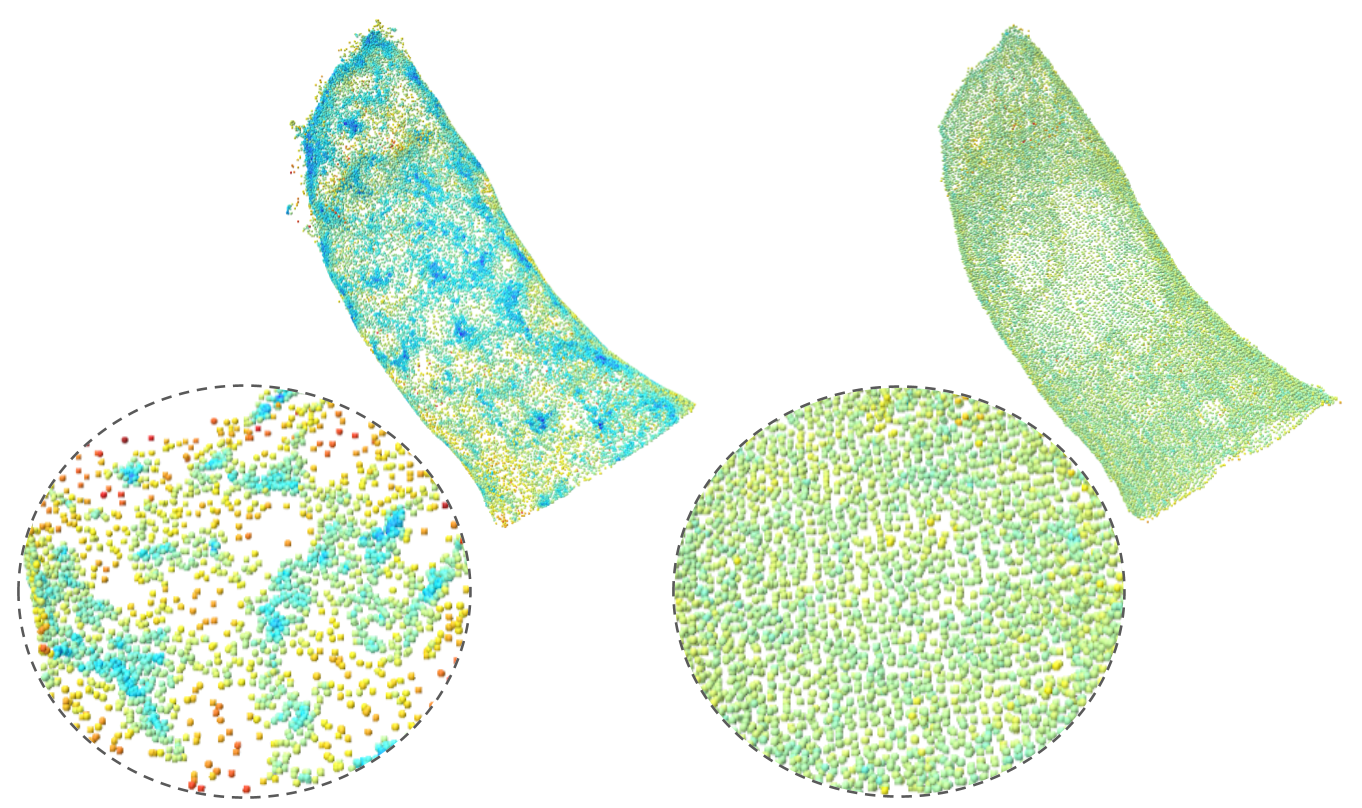}
    \put(30,  \pl){\textcolor{black}{Input}}
    \put(70, \pl){\textcolor{black}{Consolidated}}
    \end{overpic}
    \caption{\gm{Real 3D scans obtained from a low end RealSense scanner with many sparse and missing regions. Observe how the consolidated point cloud contains a more uniform sampling and eliminates noise. }}
    \label{fig:density_scanned}
\end{figure}
\subsubsection{Real Scans}
Our method was also tested on real scans from an Intel RealSense SR300 scanner, with very few points: 15K-30K. Qualitative results can be seen in Figure~\ref{fig:density_scanned}. Observe how points are up-sampled in sparse regions which results in a more uniform density, and outliers are removed in the process.
Mesh reconstruction of the last example in Figure~\ref{fig:density_scanned} can be found in the supplementary.

\begin{figure}[h!]
    \centering
    \newcommand{\pl}{-4}

    \begin{overpic}[width=\columnwidth, trim=0 0 0 0 ]{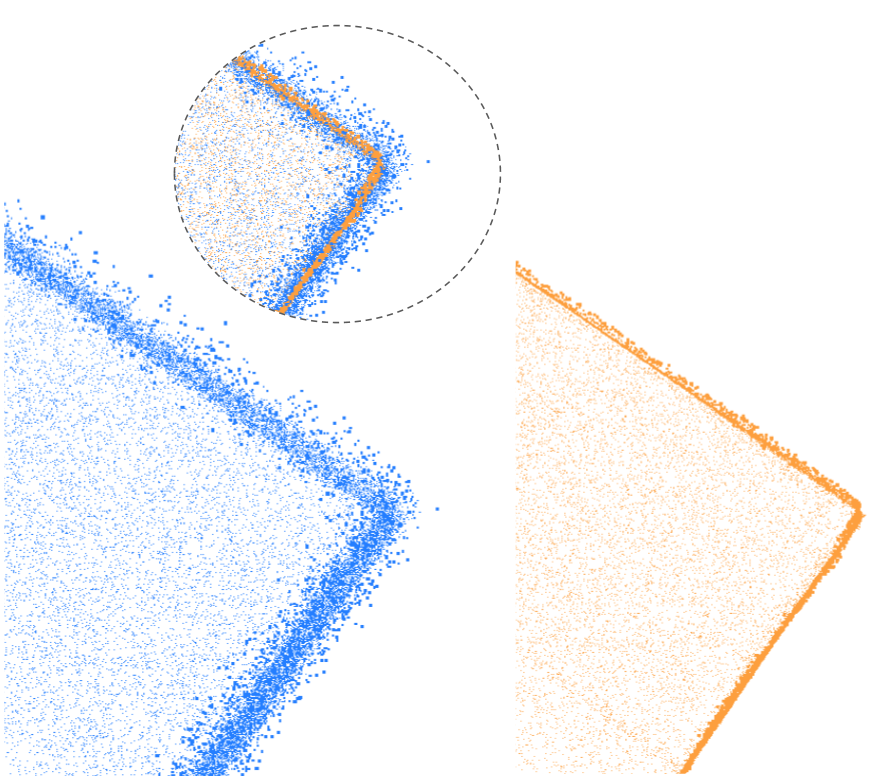}
    \put(20,  \pl){\textcolor{black}{Input}}
    \put(85, \pl){\textcolor{black}{Ours}}
    \end{overpic}
    \caption{\gm{Visual denoising results trained on a uniformly noised cube with random subsets. Observe how our method is able to reduce the noise significantly without over smoothing the edge.}}
    \label{fig:cube_denoising}
\end{figure}
\subsection{Denoising}
Although our technique is not a denoising framework per se, we obtain denoised results as a byproduct of working on global subsets. Each subset contains different samples of the noise, therefore, by training a model to match between them with an unbiased loss, the noise must be averaged out in order to minimize the loss. 
This claim is illustrated in Figure~\ref{fig:cube_denoising}. Even when training the network with random subsets \textit{i.e.,} not balanced, the generator is forced to average out the noise in order to properly minimize the training loss. Leading to a denoised result that still reliably captures the underlying surface. 

\begin{figure}[h]
    \centering
    \includegraphics[width=\columnwidth]{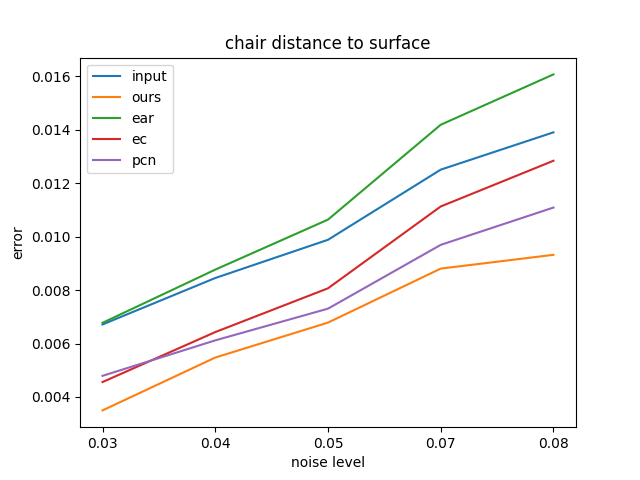}
    
    \includegraphics[width=\columnwidth]{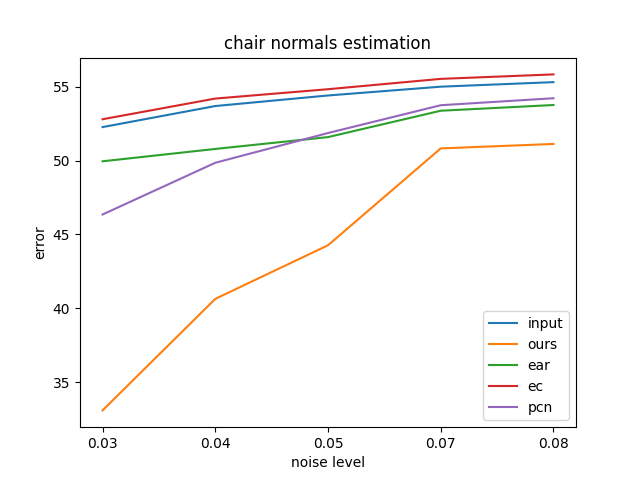}

    \caption{Distance to surface \rev{(top)} and normals estimation error \rev{(bottom)} \rev{for increasing noise levels (standard deviation)} comparison against~\cite{rakotosaona2019pointcleannet}  \cite{huang2013edge} \cite{yu2018ec}, for the chair shape present in Table~\ref{table:quant}. Observe how even a mild improvement in the distance to surface error in the low noise levels, results in significantly improved normal estimation.}
    \label{fig:denoising_graphs}
\end{figure}
To test this claim we noised the point clouds used in Table~\ref{table:quant} to varying extents and trained our method without any specific balancing, and compared the results against~\cite{rakotosaona2019pointcleannet}, \cite{huang2013edge} and \cite{yu2018ec}.
Figure~\ref{fig:denoising_graphs} shows the distance to the ground truth surface, as well as the normal estimation error compared to the ground truth normals. As expected, the distance to the ground truth surface becomes greater the more noise is added to the ground truth point cloud. Our method is able to out perform the other methods at every noise level, and its effect on the application of normal estimation is quite dramatic. The rest of the graphs for the different shapes can be found in the supplementary. 

\subsection{Implementation Details}
Our implementation uses the following software: PyTorch~\cite{paszke2017automatic}, FAISS~\cite{JDH17} and PointNet++, PyTorch~\cite{pytorchpointnetpp} and Polyscope~\cite{polyscope} for visualization.
\rev{
Training times for our method can vary depending on the shape, and is mostly affected by the amount of points and the complexity of the shape. 
The supplementary material contains a study on the convergence of the network, proposing a method to measure the \textit{novelty} of the newly synthesized points, that can also serve as a dynamic stop criterion. 
The study indicates that training takes at most roughly 1 hour and 30 minutes, but desirable results can be obtained in around 10-20 minutes. 
Interestingly, our network did not overfit, and we believe this is due to the infinitely large number of subset pairs that can possibly be sampled from the input.
}

We use the default hyper-parameters provided by the PointNet++ Pytorch implementation~\cite{pytorchpointnetpp} for the segmentation task, \gm{with six hidden layers and ReLU activations. Trained with the Adam optimizer and a learning rate of $10^{-4}$}. We selected the subset size in each iteration to be around $5-10\%$ of the input point cloud. In general, the network is insensitive to the particular subset size. An input point cloud is considered sparse if it has less than 20k points (much smaller than the amount of points typically obtained in a real scan). In this case, there may not be enough sharp points present in $5-10\%$ of the data. To this end, $5-10\%$ subsets on low resolution point clouds should be sampled uniformly instead of with curvature (see such results in the supplemental material). \rev{For EC-Net~\cite{yu2018ec} we used pre-trained weights provided from their Github repository, which was trained on manually annotated edges from 3D repositories such as ShapeNet~\cite{chang2015shapenet}. }

\section{Discussion and Conclusions}
We have introduced the concept of self-sampling for neural point cloud consolidation. The network is trained on the input data itself and infers novel points that densify and enhance the global geometric coherency of the point cloud according to the desired consolidation criterion. We have shown that with a simple formulation, our neural network achieves \rh{compelling} results, which outperforms both classic techniques and large dataset-driven learning. \rh{The proposed technique} is effective at eliminating noise and outliers, a notoriously difficult problem in point cloud consolidation.

\textbf{Global Consistency.} Unlike other point upsampling methods which analyze shapes via local patches, in this work, we learn from global subsets. Leveraging the inductive bias of neural networks leads to a globally consistent solution, and the global shared weights helps prevent \textit{overfitting}. As we have shown, the shared weights learns common local geometries across the whole shape, which allows amending erroneous regions, beyond denoising.

\textbf{Boosting Consolidation.} We categorize points based on a rough approximation of the consolidation criterion (e.g., high curvature points are mostly sharp). The \rh{appeal} of this paradigm is that it does not require that \rh{the initial point classification is perfect, or even very accurate.}
A loose definition of the desired criteria is sufficient, since networks are good at regressing to the mean, understanding the core modes of the data and ignoring outliers. This formulation can be viewed as a variant of boosting, \dc{that is,} converting set of \textit{weak} learners (\emph{i.e.,} many pairs of source and target subsets), to create one strong consolidation network. We believe that using neural networks to boost a loose definition of a desirable attribute has more applications beyond \dc{the ones presented here}. 

\textbf{Implicit Definition.} Note that our network does not learn by training with desired consolidation objective directly, but rather \textit{implicitly} through re-balanced subsets. Learning a mapping from one set of points to another, without an explicit definition of the metric used to place points in the target subset, enables the network to generalize beyond an explicit definition. There is no one-to-one mapping between source and target subsets, but rather (practically) infinitely many possible target subsets that could be selected for the loss computation. As a result, the best chance the network has to minimize the loss is by learning to generate points on the underlying shape surface with the desired consolidation criterion.

\textbf{Balancing.} The focus of this work is consolidating point clouds with an emphasis on sharp feature points or re-sampled points in sparsely sampled regions. It should be noted that these features are typically sparse, which is problematic for neural networks since they tend to ignore sparse data and focus on learning the core data modes. In our work, we compensate for sparsity by using a biased sub-sampler that re-balances the probability of sparse regions in the target subset. Note that our method does not require parameter tuning, and is generally robust to hyper-parameters.

A limitation of this approach is the run-time during inference. This technique is significantly slower than a pre-trained network or classic techniques. As learning on irregular structures becomes more mainstream and optimized (\emph{e.g.,} recently released packages Kaolin~\cite{kaolin2019arxiv} and PyTorch3D~\cite{ravi2020pytorch3d}), runtime will improve significantly.

In the future, we would like to consider extending the method to learn local geometric motifs, for example, on decorative furniture or other artistic fixtures. Learning geometric motifs can be challenging since they typically have low prevalence in a single shape, and are often unique across various shapes. Another interesting avenue is learning to transfer sharp features between shapes, possibly between different modalities, \emph{e.g.,} from 3D meshes to scanned point clouds or vice versa. We are also considering using the self-prior for increasing self-similarities within a single shape, as means to enhance the aesthetics of geometric objects, \rev{or for shape completion}.

\begin{acks}
We thank Shihao Wu for his helpful suggestions, and the anonymous reviewers for their constructive comments.
We also thank Guy Yoffe from the Technion Institute of Technology and Haifa3D for providing us with real point cloud scans.
This work is supported by the European research council (ERC-StG 757497 PI Giryes), and the Israel Science Foundation (grants no. 2366/16 and 2492/20). This work was also partially supported by the Deutsch family foundation and the Yandex initiative for machine learning.
\end{acks}

\bibliographystyle{ACM-Reference-Format}
\bibliography{bibs}

\end{document}